\documentclass[10pt,a4paper,twoside]{article}
\usepackage {graphicx} 
\usepackage{baltlat5}
\usepackage{wrapfig}
\def\apj{ApJ\,}
\def\apjl{ApJ\,}
\def\aap{A\&A\,}
\def\mnras{MNRAS\,}
\def\solphys{Sol.~Phys.\,}
\def\planss{Planet.~Space~Sci.}
\def\apss{Astrophysics and Space Science}
\def\pasp{PASP}
\def\JPG{J. Phys. G\,}

\def\pre{Phys.~Rev.~E}
\def\physa{Physica A}

\begin{document}
\ \
\vspace{-0.5mm}

\setcounter{page}{1}
\vspace{-2mm}

\titlehead{Baltic Astronomy, vol.\ts 14, xxx--xxx, 2005.}

\titleb{X-rays profiles    in symmetric 
and asymmetric supernova remnants}

\begin{authorl}
\authorb{L. Zaninetti}{1}
\end{authorl}

\begin{addressl}
\addressb{1}{
Dipartimento di Fisica Generale, Via Pietro Giuria 1,\\
           10125 Torino, Italy \\
           E-mail: zaninetti@ph.unito.it
}
\end{addressl}

\submitb{Received 2005 April 10; revised 2005 June 14}

\begin{summary}
The non-thermal X-rays from the SN 1006 NE rim
present characteristic scale lengths 
that  are interpreted in the context of diffusion
of a relativistic electron.
The adopted theoretical  framework is   the
mathematical diffusion in 3D, 1D and 1D with drift
as well as the Monte Carlo random walk in 1D with drift.
The asymmetric random walk  with diffusion
from a plane  can explain the scale widths of 0.04 pc upstream
and 0.2 pc downstream 
in the non thermal intensity   of X-ray  emission
in SN~1006.
A mathematical image of the  
non thermal  X-flux   from an   
supernova remnant
 as well
as profiles function of the distance from the center
can be simulated.
This  model provides a reasonable  description of
both  the limbs and the central region  of SN~1006.
A new  method  to deduce the magnetic field in supernova remnant
 is suggested.
\end{summary}



\resthead{X SNR}{L. Zaninetti}

\sectionb {1}{Introduction }

In order to explain the non-thermal emission
from  astrophysical objects  and the energy distribution
of cosmic rays,  the theory  of particle acceleration
was  developed using various types of acceleration mechanisms,
of  which some are  briefly reviewed as follows~:
\begin {itemize}
\item acceleration from Interstellar Medium ( in the following ISM)
     irregularities,
         Fermi (1949)
\item magnetic pumping , 
       Parker (1958)
\item 
 strong shocks ,
       Krymskii (1977)       
,
       Axford et~al. (1978)
,
       Blandford \& Ostriker (1978)
,
       Bell (1978a)        
,
       Bell  (1978b)
and
the detailed review of
       Drury (1983)
.
\item scattering by hydro-magnetic waves ,
       Skilling (1975).
\end {itemize}
At the same time, the transport equation
for the electron
energy was  set up , see for example 
       Parker (1965)
,
       Kirk (1994)
       ,
       Jokipii (1987)
         and
       Berezinskii  et~al.      (1990)
.

Further on, recent developments in the quasi-linear theory
of particle  acceleration  by super-Alfvenic shock waves
(see for example 
       Vainio  \& Schlickeiser  (1998)
,
        Klepach  et~al. (2000)
and 
       Schlickeiser (2002)
for a general review)
produced detailed
results on the spatial diffusion coefficient
but  did not   cover  the possibility of  simulating
the   X-contours of   the astrophysical object.
 We briefly review the
 effects on electron
 transport that are due to the dynamic nature of
 a scatterer , see for example  
        Skilling  (1975)
 ,equation~(9):
\begin{itemize}
 \item convection of electrons with the bulk flow of the plasma
 \item adiabatic cooling/heating of electrons in regions of
   diverging/converging flow
 \item  stochastic acceleration of electrons
\end{itemize}
 While the last effect may be considered small under some
 relatively
 reasonable restrictions, the first and  second effects
 are important
 ingredients in  the standard transport equation of electrons.
 The effect
 of convection would be to cause asymmetry in the modelled
 X--contours  by 
 pressing them closer together on the upstream side
 and further away from
 each other on the downstream side of the source.
 But these  two effects require the residence time
 of  relativistic electrons in the
 emitting   layer  to  be greater than the adiabatic/convection
 time : if  this inequality is not verified these two
 effects are negligible.

 The Monte
 Carlo modelling of particle acceleration and transport
 is not a new concept  and
 has  been extensively used in supernova remnant 
(in the following SNR) , in particular:
 \begin{itemize}
 \item
 Calculations of test particle spectra
 and acceleration times are analysed  in 
        Ellison    et~al. (1990)
  by adopting a
 first-order Fermi particle acceleration
 for parallel shocks
 with arbitrary flow velocities and compression ratios
 r up to seven, shock
 velocities u1 up to 0.98c, and injection energies ranging from thermal to
 highly super-thermal.

 \item
 Calculations concerning the  ion and electron spectra produced
 by Fermi acceleration in a steady state plane parallel modified shock for
 Mach numbers of 170 and 43  are reported in 
        Ellison \& Reynolds (1991)
 .

\item
An  extended  simple model of nonlinear diffusive shock
acceleration 
was  developed by 
       Ellison  et~al. (2000)
and 
includes the injection and
acceleration of electrons,
the production of
photons from bremsstrahlung, synchrotron, inverse-Compton, and pion-decay
processes. 
\item
X-ray   emission in the
supernova remnant G347.3-0.5
is explained by an
electron population generated by diffusive shock
acceleration at the remnant
forward shock,
see  
       Ellison et~al. (2001)
.
\end{itemize}

In astrophysics we see synchrotron emission 
from places other than the sites of active acceleration.
It is therefore a question of electron transport and in this 
paper   the following questions are   posed:
\begin {itemize}

\item   The trajectory of the relativistic electron that
        produces synchrotron radiation is helicoidal:
        on  which scales can   the trajectory
        be  approximated by  a rectilinear motion?

\item   How  can  a theory  of a particle that
        diffuses  through the random walk
        be developed
        when the mathematical diffusion is adopted?

\item  The relativistic electrons alone produce e.m.
       radiation
       at a given observing frequency
       but the spatial gradients that characterise
       the  X--sources require another treatment.
       Could a theory that  simulates the contours of
       non-thermal X--emission be set up?

\item  Can a theory of diffusion from  a plane in
       the presence of  convection  be set up?

\item The X--contours of many SNR (for example Cas~A,
      Tycho and SN~1006) present
      a ring enhancement  of  the intensity
      toward the external zones:
      does  the observed ring  have a theoretical
      explanation and can it be simulated?

\end {itemize}

The trajectories   adopted  in diffusing away from
the center of the box  were
described in Sect.~2.3.

The   mathematical diffusion   with  and without 
the convection
was  set up in Sect.~3.

The physical  bases of the random walk 
for relativistic electrons was introduced in Sect.~4.

The Monte Carlo experiment that simulates the diffusion 
of relativistic electrons from a plane  
in the presence of a stationary state situation
was  carried  out  in Sect.~5.

The formula that allows  the theoretical
counterpart of the intensity of radiation
to be built
was developed
in Sect.~6;  this mathematical 
intensity explains the observed   X-flux enhancement
in the ring  of  many SNR.

The two asymmetries in the observed flux  are  explained
with an energy cascade function of   the field 
of radial velocities  among the various regions 
of the SNR, see  Sect.~6.6.

\sectionb {2}{Synchrotron X-ray emission in astrophysical objects}

When  X-ray observations from satellites like Einstein,Chandra and
ASCA  are  directed toward typical astrophysical objects like SNR ,
super-bubbles and extra-galactic radio--sources,
they  detect non thermal
emission
with intensity I($\nu$)
$\propto$ $\nu^{-\alpha}$ (here $\nu$ is the frequency
and $\alpha$ the power law index),
see for example
       Pacholczyk (1970)
.
 The SNR, in particular, presents sharp boundaries in the
observed flux that can be explained
by invoking drift effects.
On adopting standard values of the
magnetic field,  the gyro-radius of
emitting regions  turns out to
 be $\approx$ 200 times smaller than  the typical  dimensions
 where the emission is thought to originate.
 This fact allows us to build a theory of
electron transport
 with  a  step  length 2 decades smaller than the source dimension.

These observational effects offer a starting point
with which 
\begin{itemize}
\item 
to analyse
the random walk performed with steps of  a 
physical dimension
$\frac{side(pc)}{(NDIM -1)}$,  where {\it side} \/represents
the dimension of the considered box and NDIM the
number of mean free paths contained in the box,
\item
the computation of the synchrotron losses at the light of
the random walk,
\item 
the setting  of the radiative transfer equation,
\item 
to group all the data on SNR in a unique section.
\end{itemize}

\subsectionb {2.1} {Synchrotron emission and losses }

Once the relativistic electrons are accelerated (see
Appendix~A for a brief review
on the acceleration mechanism),
\label{synchro}
the synchrotron losses  due to the magnetic field are:
\begin {equation}
{{dE}\over {dt}} =
-b_{rad} \;  E^2 \quad
{\mathrm{{erg\; sec^{-1}}}}  \quad,
\label{eq:losses}
\end {equation}
in cgs  $E$ is expressed in \mbox {ergs} ,  $b_{rad}$
=  $1.579\times 10^{-3}H^2 $ and  H is expressed in Gauss.
Another type of loss  is  the adiabatic one,
but in Appendix~B
we have  shown that they  can be neglected
because  the residence  times of the
relativistic electrons are  small.

We will see  in Sect.~4.2.2 that the
synchrotron  losses  are negligible in the
X-region and consequently the working hypothesis
of the random walk without losses is justified.

\subsectionb {2.2} {Radiative transfer equation}

The transfer equation in the presence of emission only
, see for example  
       Rybicki \&  Lightman  (1985)
 or
       Hjellming (1988)
 ,
 is
 \begin{equation}
\frac {dI_{\nu}}{ds} =  -k_{\nu} \zeta I_{\nu}  + j_{\nu} \zeta
\label{equazionetrasfer}
\quad ,
\end {equation}
where  $I_{\nu}$ is the specific intensity , $s$  is the
line of sight , $j_{\nu}$ the emission coefficient,
$k_{\nu}$   a mass absorption coefficient,
$\zeta$ the  mass density at position s
and the index $\nu$ denotes the interested frequency of
emission.
The solution of equation~(\ref{equazionetrasfer})
 is
\begin{equation}
 I_{\nu} (\tau_{\nu}) =
\frac {j_{\nu}}{k_{\nu}} ( 1 - e ^{-\tau_{\nu}(s)} )
\quad  ,
\label{eqn_transfer}
\end {equation}
where $\tau_{\nu}$ is the optical depth at frequency $\nu$
\begin{equation}
d \tau_{\nu} = k_{\nu} \zeta ds
\quad.
\end {equation}
The synchrotron emission  values of
$j_{\nu} \zeta$ and $k_{\nu} \zeta$ can be found in
       Hjellming (1988)
.
We now continue analysing the case of optically thin layer
in which $\tau_{\nu}$ is very small
( or $k_{\nu}$  very small )
and the density  $\zeta$ is substituted
with our concentration C(s) of relativistic electrons
\begin{equation}
j_{\nu} \zeta =K_e  C(s)
\quad  ,
\end{equation}
where $K_e$ is a  constant function
of the energy power law index,
the magnetic field
and the frequency of e.m.\ emission.
The intensity is now
\begin{equation}
 I_{\nu} (s) = K_e
\int_{s_0}^s   C (s\prime) ds\prime \quad  \mbox {optically thin layer}
\quad.
\label{transport}
\end {equation}
The increase in brightness
is proportional to the concentration integrated along
the line of  sight.
In  Monte Carlo experiments
the concentration is memorised  on
the   grid
${\mathcal M}$ and the intensity is
\begin{equation}
{\it I}\/(i,j) = \sum_k  \triangle\,s \times  {\mathcal M(i,j,k)}
\quad  \mbox {optically thin layer}\quad, 
\label{thin}
\end{equation}
where $\triangle$s is the spatial interval between
the various values and  the sum is performed
over the   interval of existence of the index $k$.
The theoretical flux density is then obtained by integrating
the intensity at a given frequency  over the solid angle of the
source.
In order to deal with the transition to the optically thick case
in the Monte Carlo simulation , see Sect.6.2,
the intensity is given
by
\begin{equation}
{\it I}\/(i,j) = \frac {1}{K_a}  (1 - \exp (- K_a\sum_k  \triangle\,s \times  {\mathcal
M(i,j,k)}))
\quad  \mbox {thin $\longmapsto $ thick }\quad,
\label{transition}
\end{equation}
where  $K_a$ is a constant that represents the absorption.
Performing a  Taylor expansion of  the last formula (\ref{transition}),
the equation~(\ref{thin})  is obtained.

\subsectionb {2.3} {The  trajectories }

\label{traiettorie}

Here the  approximation of trajectories moving on a
straight line is used
and the  randomisation on  the two or six 
main directions is  due to the presence of scattering
clouds (without re-acceleration) ,
see Appendix~A.
The  length of the step  $\delta$  is  now compared
with  the length where the standard theories assume that
the transport is working.
If the magnetic field is strongly turbulent,
the mean
free path of an electron can  be identified
with its gyro--radius.
The gyro--radius of the electron can be parametrised
 by
using  the  synchrotron emission frequency 
expressed in MHz  units , $\nu_{M}$ , and the magnetic
field expressed  in $10^{-4}$ Gauss units, $H_{-4}$:
\begin  {equation}
\rho = 2.6\;10^{-10}
\frac {(\nu_{M})^{1/2}}
      {(H_{-4})^{3/2}}  pc
\label{gyroradius}
\quad.
\end    {equation}
This should be considered an approximate relationship 
since  even a single electron produces an emission spectrum of finite
width.
On the insertion of  the magnetic field  of SN~1006,
$H_{-4}$ = 0.15 (see discussion on the data of
SN~1006 in Sect.~2.4) 
  the value  of the gyro-radius
now  depends exclusively  on the chosen   emission frequency.
A typical  frequency  belonging to the radio--window
is 1~GHz
which corresponds to  $\rho=0.141~10^{-6}$~pc.
The   thickness of the emitting shell is    $0.52$~pc
which means  that NDIM should be  $3.6~10^6$
once we assume that the length of the   step
is equal to the
gyro-radius.

Another zone of the e.m.\ spectrum  analysed here is
X-ray  emission, see Sect.~6.
In this case the frequency observed
is  $4~10^{17}~Hz$ which means  that $\rho=2.8~10^{-3}$~pc
and NDIM=184.

The transport of  relativistic electrons 
with the length of the step
equal to the relativistic electron  gyro-radius is  
called Bohm diffusion 
      Bohm et~al. (1949)
and   the diffusion coefficient will be frequency (energy) dependent.
The assumption of the Bohm diffusion allows to fix a one-to-one 
correspondence between frequency (energy)  and length of the step
of the random walk.

\begin{figure*}
\begin{center}
\includegraphics[width=12cm,angle=-90] {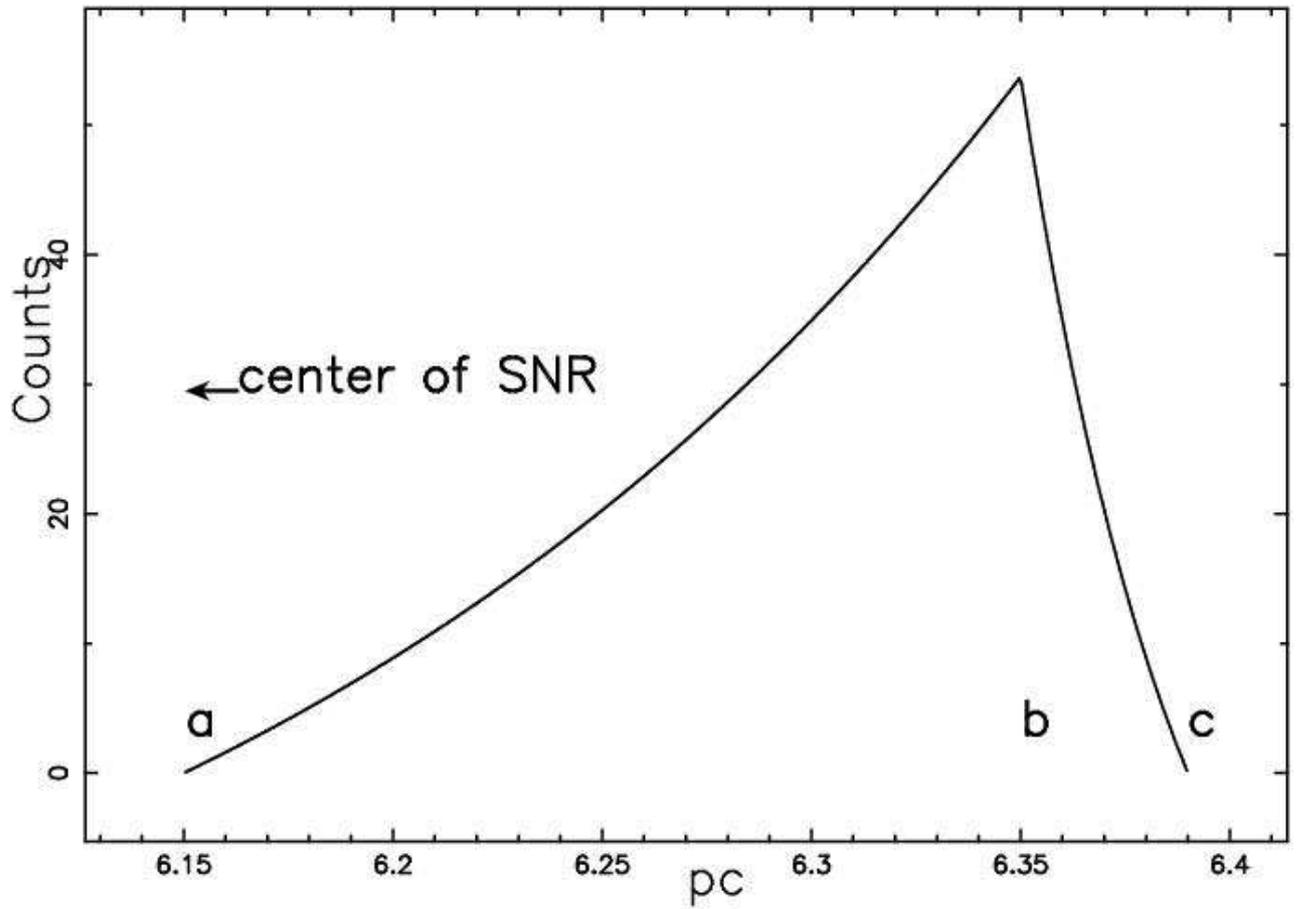}
\end {center}
\caption
{
Profile of intensity of a non-thermal filament in the SN 1006 northeast 
shell in the X-band inserted in the small box.
Adapted from Figure 4  in 
       Bamba  et~al. (2003)
.
}
\label{f01}
    \end{figure*}

\subsectionb {2.4} {The data on SN1006}
\label{data_1006}
The diameter of the known remnants  spans the range
from  3~pc
to 60~pc  and    attention is fixed on SN~1006  and its
possible  diameter  of 12.7 \mbox{pc}, 
see 
       Strom (1988)
.
Information on  the thickness of emitting layers
is contained  in  a recent study by 
       Bamba  et~al. (2003)
  which  analyses
Chandra observations (i.e., synchrotron X-rays) from SN~1006.
The observations found
that  sources of non-thermal radiation are likely to be
thin sheets
with a  thickness of about 0.04 pc
upstream  and 0.2 pc
downstream of
the surface of maximum emission,
which coincide with the locations of
Balmer-line optical
emission , see 
       Ellison et~al. (1994)
.
The  values of $W_u$ and   $W_d$ are 
now introduced 
in the same way as   equation~(1)  by  
       Bamba  et~al. (2003)
(the values of x
after which  the X-ray intensity is smaller
by  a factor  {\it e}
\/in the upstream and downstream 
directions  respectively) 
\begin{equation}
f(x) = \left\{\begin {array} {ll}
      A~\exp - | \frac{x_0 - x}{W_u} | & upstream \\
      A~\exp - | \frac{x_0 - x}{W_d} | & downstream~.
                \end  {array}
                \right.
\label{exponential}
\end{equation}
Here A is the amplitude of the cut and  $x_0$ is the
position of maximum emission.
Observation  of 
the six filaments , see 
       Bamba  et~al. (2003)
, 
gives  the following averaged values in the case of non-thermal emission:
$W_u$= 0.04~pc and  $W_d$= 0.2~pc.
These observations allow us 
 to build  the intensity of an averaged
theoretical X-filament,
Figure~\ref{f01}. 

For reasons that will be clarified in Sect.~3
we introduce $b$=$\frac{SNR~diameter}{2}$ = 6.35~pc ,
                                   $a$ =$b$ -$W_d$= 6.15~pc 
                            and    $c$ =$b$ + $W_u$= 6.39~pc.
Another interesting parameter is the 
{\it side} \/of the cubical box situated
between the internal absorbing sphere of radius $a$ 
and  the external absorbing sphere   of radius  $c$ ;
this small box will be useful to simulate the transport of particles 
from the shock region toward the upstream and downstream 
regions.

The thickness of the layer 
, $\Delta~R$ ,
 is    $\frac{SNR~radius}{12}$
according to 
       McCray  (1987)
 and allows us to deduce 
the  theoretical  value {\it side}=0.52~pc.
Another useful cubical box   includes all 
the SNR and has , for example , $ side_{SNR}$=18.37 pc ;
this will be called the great box.
The value here adopted in  the magnetic field , $H_4$=0.15,
is similar to $H=10\mu~Gauss$  of
        Bamba  et~al. (2003)
  ;
        Dyer et~al. (2004)
  quote  a smaller
value ,   $H=4~\mu Gauss$~.
The data  presented here plus the spectral index
on non thermal X-radiation (see 
       Bamba  et~al. (2003)
)
plus the ratio
of flux in the ring/center regions
,
       Dyer et~al. (2004)
, are reported in Table~\ref{data_snr} 
together the meaning of the symbols.

 \begin{table} 
 \caption[]{The data on SN~1006 } 
 \label{data_snr} 
 \[ 
 \begin{array}{llc} 
 \hline 
 \hline 
 \noalign{\smallskip} 
 symbol  & meaning & value  \\ 
 \noalign{\smallskip} 
 \hline 
 \noalign{\smallskip}
a  & radius~of~the~internal~absorbing~sphere & 6.15~pc \\ \noalign{\smallskip}
b  & radius~of~the~shock                        & 6.35~pc  \\ \noalign{\smallskip}
c  & radius~of~the~external~absorbing~sphere & 6.39~pc \\ \noalign{\smallskip}
side  & theoretical~distance~between~internal  &0.52~pc   \\ \noalign{\smallskip}
~& and~external~absorbing~sphere &~\\ \noalign{\smallskip}
side_{SNR}   & side~of~a~big~box~&  18.37 pc \\  \noalign{\smallskip}
~& including~the~SNR &~\\  \noalign{\smallskip}
H_{4}  & magnetic~field~in~10^{-4}~Gauss & 0.15 \\ \noalign{\smallskip}
u_s    & shock~velocity~in~km~s^{-1}    & 2600 \\ \noalign{\smallskip}
\Gamma   & spectral~index (2.0-10.0 keV) & (2.1-2.4) \\ \noalign{\smallskip}
\frac {NT~flux~limb} {NT~flux~center} & ratio~of~X-NT~flux~at~limb  &
(5.4-10.8)       \\ \noalign{\smallskip} 
\noalign{\smallskip} 
~& on~X-NT~flux~at~center  &~        \\ \noalign{\smallskip} 
\noalign{\smallskip} 

 \hline 
 \hline 
 \end{array} 
 \] 
 \end {table} 
One should remember that
the profiles of
non-thermal X-ray  emission
in 
       Bamba  et~al. (2003)
 are found in another diffusion model.
They are extracted from the total X-ray flux when the thermal
component is not negligible. Suggestions for a  new diffusion
model will be  presented in Sect.~3  but up to
that section the exponential profiles in
       Bamba  et~al. (2003)
 will be
the observational reference. Roughly speaking we can say that
given two decreasing profiles  of intensity , downstream and
upstream in respect to the position of the advancing shock , the
exponential fits can represent a first approximate description of
a more complex behaviour of the intensity , see
Sect.~3.

\sectionb {3} {Mathematical diffusion}

\label{mathematical}
Once the concentration , $C$,
the energy per electron or nucleon , $\epsilon_k$ ,
the spatial coordinates, $\bf{r}$     ,
the diffusion tensor     $\widehat{D}(r,\epsilon_k)$
are introduced
and the losses are neglected
the  general diffusion  transport equation for a specie
(electrons or nucleons)  is
 ,  see for example equation~(3.1) in
       Berezinskii  et~al. (1990)
,
\begin{equation}
\frac {\partial C }{\partial t} =
\nabla \cdot (\widehat{D}(r,\epsilon_k)) \nabla C
\quad.
\label{eqndiffusion}
\end{equation}
In general  $\widehat{D}(r,\epsilon_k)$   is variable
 both for
the standard "thermal" diffusion and for the Bohm one. Here we
will concentrate on the case of constant energy ( or negligible
losses ) that in our astrophysical case means a given frequency of
observation , see formula~(\ref{ro}). Therefore
$\widehat{D}(r,\epsilon_k)$ is independent 
from the spatial coordinates and transforms in D that is now constant;
of course the given energy/frequency of astrophysical interest
should always be specified because the mean free path (i.e. Bohm
diffusion ) changes. The diffusion equation~(\ref{eqndiffusion})
becomes the Fick'~s second equation ,  see for example
      Berg  (1993)
, and it's expression is different according to the
chosen dimension. In three dimensions it is
\begin{equation}
\frac {\partial C }{\partial t} =
D \nabla^2 C
\quad,
\label{eqfick_3}
\end {equation}
in one dimension   it is
\begin{equation}
\frac {\partial C }{\partial t} =
D  \frac {\partial^2C}{\partial x^2}
\quad.
\label{eqfick_1}
\end {equation}

In one dimension  and in the presence of a drift velocity
along the x-direction $u$
 is 
\begin{equation}
\frac {\partial C }{\partial t} =
D  \frac {\partial^2C}{\partial x^2} -  {\vec{u}} \frac {\partial C}{\partial x}
\quad.
\label{eqfick_1_drift}
\end {equation}
The hypothesis of the steady state allows us to deduce simple solutions
for the concentration in 3D , 1D and 1D with drift
and for the spectral index.

The physical justification  that allows to set the concentration 
equal to zero at a given distance from the source   is reported in 
Sect.~4.2.3  and  Sect.~6.3.

The solutions for  the diffusion equation 
can  also be obtained 
with the usual
methods of eigenfunction expansion 
(see 
       Gustafson (1980)

 and   
      Morse  \& Feshbachm (1953)
)  and 
they are  cosine  series 
whose coefficients decay exponentially;
they  can be deduced from formula~(25) 
of 
       Ferraro  \& Zaninetti (2004)
 upon inserting 
the  dimension $d$.

\subsectionb {3.1} {3D case}

\begin{figure*}
\begin{center}
\includegraphics[width=12cm,angle=-90]{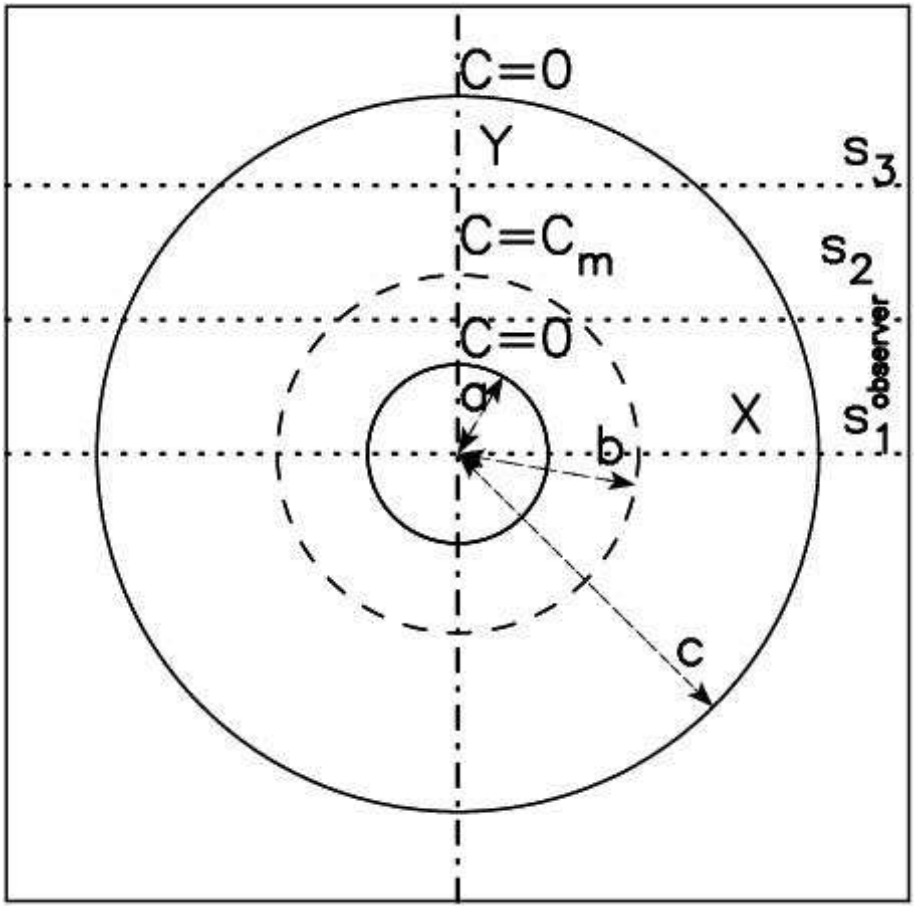}
\end {center}
\caption
{
The spherical source inserted in the great  box is  represented through
a dashed line, and the two absorbing boundaries
with a full line.
The observer is situated along the x direction, and 
three lines of sight are indicated.
Adapted from Figure~3.1 by 
       Berg  (1993)
.
}
\label{f02}
    \end{figure*}

Figure~\ref{f02} shows 
a spherical shell  source of radius  $b$
between a spherical absorber
of radius $a$ and a spherical absorber of radius $c$.

\begin{figure*}
\begin{center}
\includegraphics[width=12cm,angle=-90]{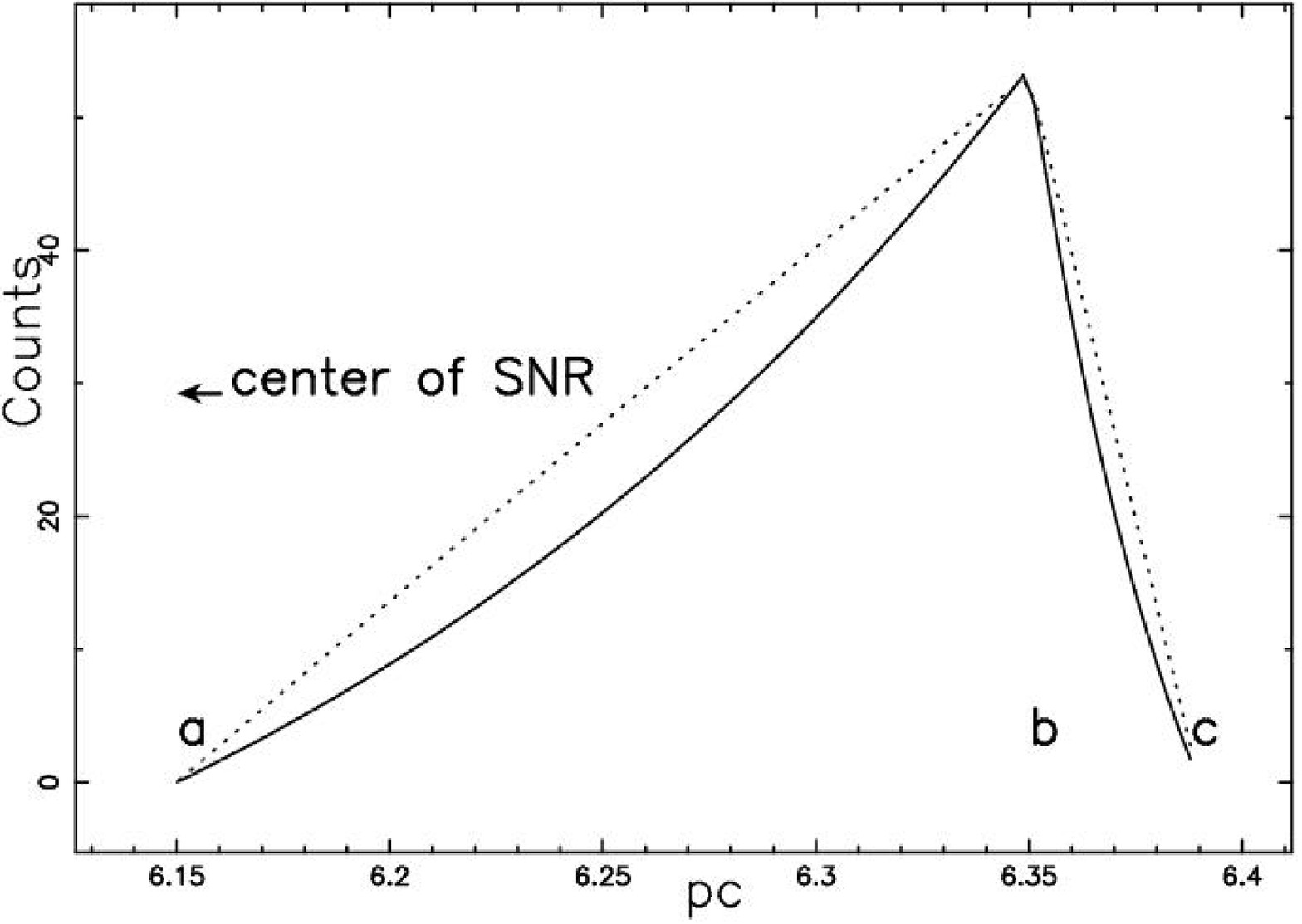}
\end {center}
\caption
{
Observed profile of the  non-thermal X filament in SN 1006 
(full line ) and  spherical solution of the 3D concentration
 (dotted line)
inserted in the small box.
Parameters as  in Table~1.
}
\label{f03}
    \end{figure*}

The  concentration raises from 0 at {\it r=a}  to a
maximum value $C_m$ at {\it r=b} and then  falls again
to 0 at {\it  r=c}~.
The  solution of  equation~(\ref{eqfick_3})
in presence of steady state is
\begin{equation}
C(r) = A +\frac {B}{r}
\quad,
\label{solution}
\end {equation}
where $A$ and $B$  are determined by  the boundary conditions~,
\begin{equation}
C_{ab}(r) =
C_{{m}} \left( 1-{\frac {a}{r}} \right)  \left( 1-{\frac {a}{b}}
 \right) ^{-1}
\quad a \leq r \leq b
\quad,
\label{cab} 
\end{equation}
and
\begin{equation}
C_{bc}(r)=
C_{{m}} \left( {\frac {c}{r}}-1 \right)  \left( {\frac {c}{b}}-1
 \right) ^{-1}
\quad b \leq r \leq c
\quad.
\label{cbc}
\end{equation}
These solutions can be found in 
       Berg  (1993) 
or in 
       Crank 1979
. 
Thus a comparison can be made  between the observed and the 
theoretical spherical solution , see Figure~\ref{f03}.

\subsectionb {3.2} {1D case}

\begin{figure*}
\begin{center}
\includegraphics[width=12cm,angle=-90]{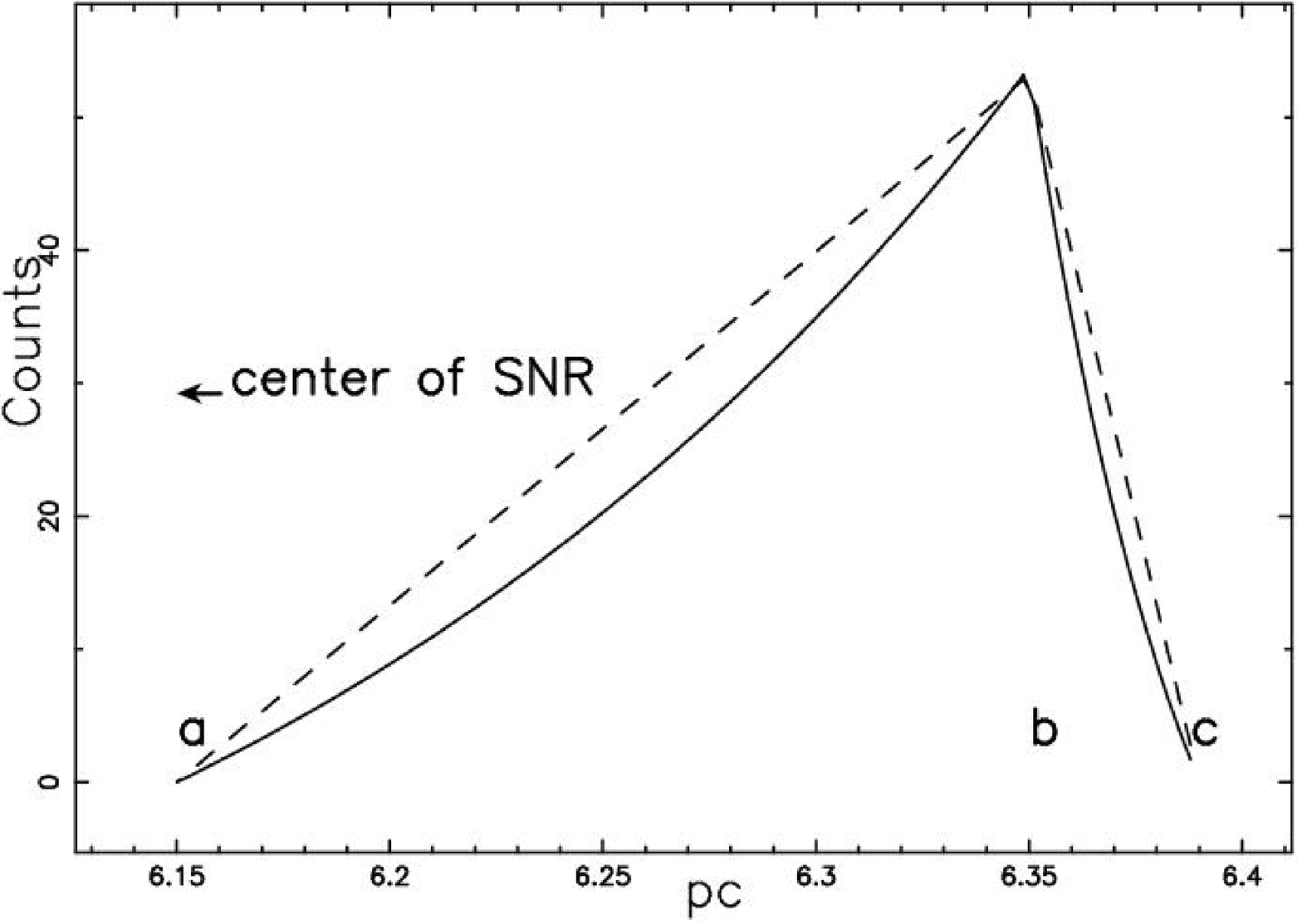}
\end {center}
\caption
{
Observed profile of the  non-thermal X filament in SN 1006 
(full line ) and  1D solution of the concentration  (dashed line)
inserted in the small box.
Parameters as  in Table~1.
}
\label{f04}
    \end{figure*}

The 1D solution of the concentration  is the same as   
the diffusion through  a plane sheet.
If we have 
 a point  source at distance  {\it  b}
between a point  absorber 
of distance {\it a} and a point absorber of distance {\it c},
the general solution of equation~(\ref{eqfick_1})
as a function of the variable x that denotes the distance
in presence of steady state is  
\begin{equation}
C(x) = A + B x 
\quad.
\label{solution_1D}
\end{equation}
The boundary condition  gives 
\begin{equation}
C(x) =
C_{{m}}  \frac {x-a}{b-a} 
\quad a \leq x \leq b
\quad, 
\label{cab_1d}
\end{equation}
and 
\begin{equation}
C(x) =
C_{{m}}   \frac {x-c}{b-c} 
\quad b \leq x \leq c
\quad.
\label{cbc_1d}
\end{equation}
Figure~\ref{f04} reports the observed X-profiles as well
as the profiles of the 1D concentration as given by 
the diffusion from a plane.

\subsectionb {3.3}{1D case with drift}

\begin{figure*}
\begin{center}
\includegraphics[width=12cm,angle=-90]{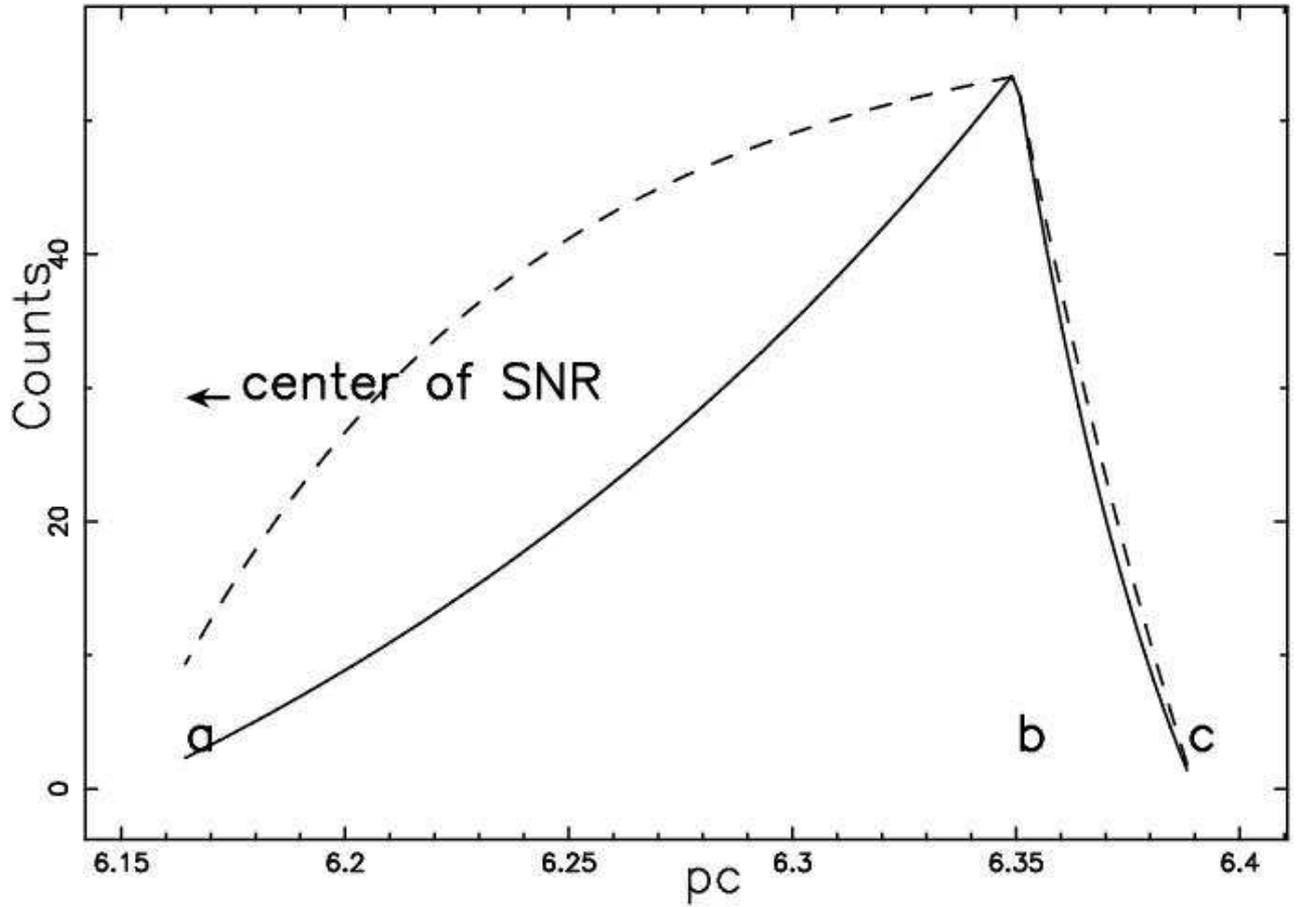}
\end {center}
\caption
{
Observed profile of the intensity of  non-thermal X filament in SN 1006 
(full line ) and  1D solution of concentration   in the presence
of drift , $\frac {u}{D}=12.2~pc^{-1}$   (dotted  line),
see formula~(\ref{formulaud}) inserted in the small box.
Parameters as  in Table~1.
}
\label{f05}
    \end{figure*}

Also here  a point  source at distance   {\it b}
between a point  absorber
at distance $a$ and a point absorber at distance $c$ are considered.
The general solution of equation~(\ref{eqfick_1_drift})
in presence of steady state is 
\begin{equation}
C(x) = A + B e^{{\frac{\vec {u}}{D}}x } 
\quad. 
\end{equation}
The advection velocity here considered is the downstream velocity
of the shock that has direction toward the center
of the SNR , conversely the triplet {\it  a,b,c,} determines a positive direction
toward the advancing shock in the laboratory frame.
In our application {\it u}  and {\it x} 
 have opposite directions and therefore 
{\it u } is negative ;  the solution is
\begin{equation}
C(x) = A + B e^{{-\frac{u}{D}}x } 
\quad,
\label{solution_1D_drift}
\end{equation}
now the velocity $u$ is a scalar.
The boundary conditions  gives 
\begin{equation}
C(x) =
\frac {C_{{m}} } {e^{{- \frac{u}{D}}b  }  -  e^{{-\frac{u}{D}}a  }}
\bigl ({e^{{-\frac{u}{D}}x  }  -  e^{{-\frac{u}{D}}a  }}\bigr )
\quad a \leq x \leq b~\quad downstream~side
\quad, 
\label{cab_drift}
\end{equation}
and 
\begin{equation}
C(x) =
\frac {C_{{m}} } {e^{{-\frac{u}{D}}b  }  -  e^{{-\frac{u}{D}}c  }}
\bigl ({e^{{-\frac{u}{D}}x  }  -  e^{{-\frac{u}{D}}c  }}\bigr )
\quad b \leq x \leq c~\quad upstream~side
\quad.
\label{cbc_drift}
\end{equation}
Figure~\ref{f05} reports the observed X-profiles of intensity
 as well
as the asymmetric  profiles of 1D concentration as given by the 
mathematical diffusion from a plane 
in the presence of drift.

\subsectionb {3.4} {Variable spectral index in 1D}

The  diffusion loss equation for relativistic electrons 
is usually solved by neglecting the spatial diffusion.
The opposite can be done by neglecting the energy losses  
and exploring how the variations in the absorbing boundaries 
influence the  spectral index.
Up to now the more interesting result on the prediction 
of the spectral index resulting from particle acceleration 
in shocks , i.e. SNR or Super-bubble , is due to 
       Bell (1978a)        
,
       Bell (1978b)        
and 
       Longair (1994)
.
The predicted differential energy spectrum of the high energy 
electrons is 
\begin{equation}
C(E)dE \propto E^{-2} dE 
\quad.
\end {equation}

But this is the spectral index 
expected  where the electrons  are accelerated and
due to the spatial diffusion the spectral index changes.  

Let us consider two energies $E_{1}$ and  $E_{2}$ ($E_2~>~E_1$) 
with two corresponding numbers  of electrons $C_1$ and $C_2$
($C_1~>~C_2$) : for example with $E_2/E_1=10$, $C_1/C_2=100$ .

\begin{figure*}
\begin{center}
\includegraphics[width=12cm,angle=-90]{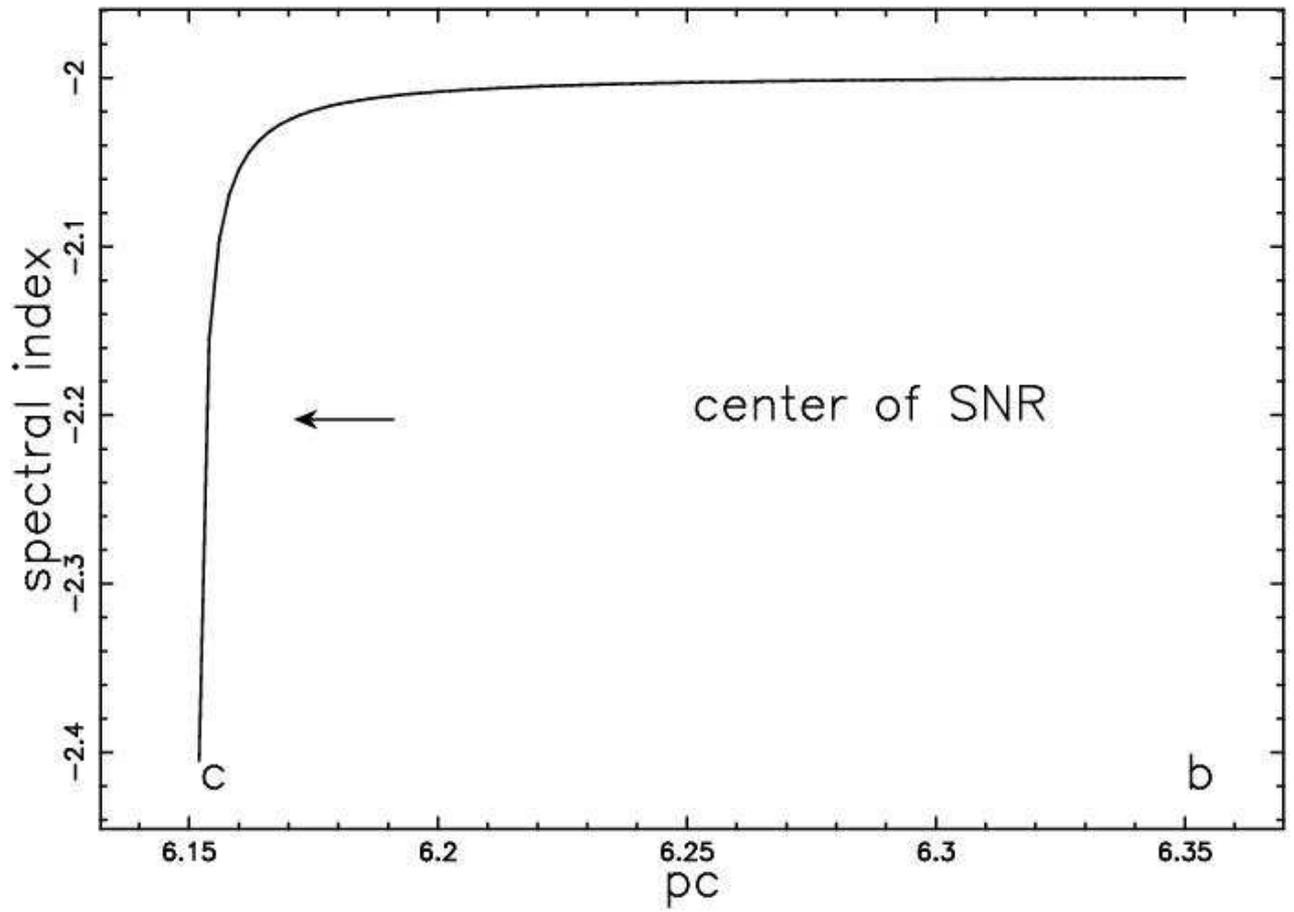}
\end {center}
\caption
{
Behaviour of the spectral index as a function 
of the distance from injection point $b$
in the downstream region of SN1006 ;
$a_1$ =a ,  $a_2=1.0002~a_1$
and other parameters as  in Table~1.
}
\label{f06}
    \end{figure*}

In  the framework of the 1D  diffusion,  the spectral index
$\Gamma$ of the energy will vary along the  distance from
the injection point according to the following formula:
\begin{equation}
\Gamma = Log_{10} \frac {C_1}{C_2}
\quad.
\end{equation}
In order to allow the spatial index to vary, it  is enough to modify
the absorbing distances of the 
two populations of electrons , $a_1$ =a and $a_2~>a_1$.
This effect is reported in Figure~\ref{f06} where 
variation  of the spectral index
cover  the observed range in the X-region ,
see Table~1.

\sectionb {4} {The physics of the diffusion}

\label{bases}
\label{sectionx}
The  reliability  of the stationary state,
evaluations on the residence times and the influence
of  synchrotron losses on the diffusion
are now analysed in  the light of Monte Carlo diffusion 
on a discrete 1D lattice.
In the following,  NDIM indicates the number of grid points 
and NPART the number of particles with different patterns.

\subsectionb{4.1}{Stationary state}

\label{sec:stationary}
It is  important to stress that our
simulations involving the 1D or 3D random walk  cover the stationary
state.  Suppose, for  example,
 that an  electron reaches the boundaries ( one of the two
absorbing boundaries  )  after n steps.
The visitation/concentration grid and the energy state are also marked
for the previous (n-1), (n-2)\ldots1 steps.
This means that the (n-1) step was  due to  the electron
emitted  one  unit of time later than that characterised
by n step,
the (n-2) step was due to the electron emitted two units of time
later  than that characterised
by n step  and  so on~.

This  procedure can be replicated   for all the n steps ; this
is equivalent to saying that  $\frac {dN}{dt}=0  $
or that
we are treating  {\it the stationary state}.

We now calculate  the time scales that the electrons spend inside the
simulation box in order to test "The stationary state hypothesis".
The time elapsed from the injection in the central
point is:
\begin{equation}
t_{rw} =
N_{step} \delta (pc) 2 /c =
N_{step} \delta (pc)~6.52~yr
\quad,
\label{nstep}
\end{equation}
where $N_{step}$  is the number of steps before
reaching  one
of the two boundaries and the time of crossing
the distance $\delta$ is doubled according to
formula~(\ref{twice}).
The number of steps  necessary to reach the  boundaries
depends on  the numbers   NDIM   and
NPART  : its distribution   is characterised by
a maximum value  $N_{step}^{max}$   which  is
extracted  from the simulation~.
The inequality characterising the "The stationary state
hypothesis"  is :
\begin{equation}
t_{exp}~>~t_{rw}
\quad,
\label{stationary}
\end{equation}
where $t_{exp}$ is defined in \ref{sec_adiabatic}.

The stationary  state hypothesis can
be checked  through  a numerical and
analytical argument. 
These two arguments    depend  fully  on the
choice of diffusion , basically $\overline{R^2}\propto N$  ; other
choices of D can produce  different results.

Here we have assumed that the acceleration times are smaller 
than the time of losses or expansion  or in other words 
the electrons are accelerated in a smaller region in respect to 
that one in which they diffuse.
A detailed discussion of the concept of the steady state 
when acceleration and cooling times are comparable 
can be found in 
       Drury (1983)
.

\subsubsectionb{4.1.1}{Numerical estimate}

In every simulation    the following
inequality should be checked:
\begin{equation}
N_{step}^{max}~<~\frac {2.5 \times t_{SNR}}{6.52
\times \delta(pc)}
\quad  .
\label{nstepmax}
\end{equation}

\subsubsectionb{4.1.2}{Analytical estimate}

A simple analytical  estimate for the residence time  of
a particle diffusing  out from a box of dimension $L$ is presented.
The typical  diffusing time $\tau$   is
\begin {equation}
\tau = \frac {\overline R^2}{2dD} \quad,
\label{meansquare}
\end{equation}
where  $\overline{R^2(t)}$ is
the mean square displacement
(see 
       Gould  \& Tobochnik (1988)
,~equation~(8.38~)) ,
$D$=$\frac {1}{2} \frac {c}{2} \delta $ is  the diffusion coefficient 
and $d$ is the dimension , in our case 1.
On assuming that   $R=L/2$ we obtain 
\begin {equation}
\tau =  \frac {1}{2} \frac {L}{\delta } \frac {L}{c}
\quad,
\end{equation}
where  the mean free path  has been replaced
by the length of the step $\delta$.
The steady state hypothesis
is represented by
\begin {equation}
\tau  \ll t_{exp} = \frac{5}{2}  t_{SNR}
\quad,
\end{equation}
which  corresponds to
\begin {equation}
NDIM   \ll t_{exp} = 5~t_{SNR}  \frac{c}{L}
\quad,
\end{equation}
once $\frac{L}{\delta}$  has been replaced by NDIM~.
Assuming  $L=\frac{R_{SNR}}{12}$, 
the inequality becomes
\begin {equation}
NDIM   \ll  60~t_{SNR}  \frac{c}{R_{SNR}} \quad.
\end{equation}
The radius of a SNR is known to be
$R_{SNR}\propto  t_{SNR}^{2/5}$
\begin {equation}
V_{SNR}= \frac{2}{5} \frac {R_{SNR}} {t_{SNR}}
\quad,
\end{equation}
and the inequality becomes
\begin {equation}
NDIM   \ll  24  \frac{c}{V_{SNR}}
\quad,
\end{equation}
and given  $\frac {c}{V_{SNR}}~\approx~100$
\begin {equation}
NDIM   \ll  2400
\quad.
\end{equation}

\subsectionb{4.2}{Residence time}
The  existence  of the random walk  modifies  the  canonical
formula that  allows  us  to compute  the distance  over which
a relativistic electron  has  damped its  energy~.
An electron that  loses its  energy  due to radiation damping
has a lifetime  $\tau_r$  ,
\begin{equation}
\tau_r  \approx  \frac{E}{P_r} \approx  500  E^{-1} H^{-2} sec
\quad  ,
\label {taur}
\end{equation}
where  $E$  is the energy in ergs  ,
H the magnetic field in Gauss,
and  $P_r$  the total radiated
power , see 
       Lang (1999)
 ,formula (1.157).
The synchrotron radiative lifetime  of an electron depends 
on pitch angle ;  thus equation~(\ref{taur}) is to 
be regarded as a typical value
of $\tau_r$ for an approximately isotropic , 
highly relativistic 
electron population.

The  energy  is connected  to  the critical  frequency,
 see 
       Lang (1999) 
,formula (1.154) as ,
\begin {equation}
\nu_c = 6.266 \times 10^{18} H E^2~Hz
\quad  .
\label {nucritical}
\end{equation}
The total power radiated in frequency interval , see 
formula (1.160) in 
       Lang (1999)
 , is given by
$P(\nu) \propto F (\nu/\nu_c)$ with
\begin{equation}
F (\nu/\nu_c) =
\int _{\nu/\nu_c} ^\infty  K_{5/3} (\eta) d\eta 
\quad  ,
\end{equation}
where $K_{5/3} (\eta)$ is a modified Bessel function.
The  function $F (\nu/\nu_c)$ has a maximum
at $\nu/\nu_c$=0.29  and some authors , see for example
       Reynolds (1998)
 , use this numerical relationship
in order to identify the energy of the electron
with  the observed frequency.
Is also interesting to compute
the   averaged frequency of emission 
$\overline{\nu}/\nu_c$  computed as 
\begin{equation}
\frac {\overline{\nu}}
      {\nu_c} = 
\frac {\int _{0.1} ^{10} (\nu/\nu_c) F (\nu/\nu_c) } 
      {\int _{0.1} ^{10}  F (\nu/\nu_c) }=1.29
\quad  .
\end{equation}
In the following we will use $\nu/\nu_c=1$ 
that can be considered an average between the two points  
of view previously expressed.

Upon inserting  in  (\ref{taur})  the energy as  given in
(\ref{nucritical})  :
\begin{equation}
\tau_r  = 1.251 \times 10^{12}  \frac {1}{\nu^{1/2} H^{3/2}}  sec
\quad  .
\label{taurnu}
\end{equation}
The  maximum frequency  that  can escape  from the center
of  the box   and  the
parameters for which  the gyro-radius  equals the
length of the step   are now computed.

\subsubsectionb{4.2.1}{Maximum frequency allowed}

If  a box with a  side $L$ has  an electron inserted
at the center,
 the classical time of crossing
the box  $t_c$  is :
\begin{equation}
t_c  =\frac{L}{2c}
\quad  .
\label {tc}
\end {equation}

The  solution  of the equation
\begin{equation}
\frac {\tau_r}{t_c} =1
\quad,
\end {equation}
gives the  maximum allowed frequency  , $\nu_c^{max}$
\begin{equation}
\nu_c^{max} =  5.962 \times10^{20} \frac {1}{L_{pc}^2 H_{-4}^3}~Hz
\quad,
\label  {nucmax}
\end{equation}
where  $L_{pc}$  is the side in pc  and  $H_{-4}$  the magnetic
field  in $10^{-4}$ Gauss   units~.
Upon  inserting the typical parameters of SN~1006  ,
reported in Table~1:
\begin{equation}
\nu_c^{max} =  6.490 \times10^{23}~Hz
\quad  .
\end{equation}
For greater frequencies
the electron cannot  escape
from the box when the classical treatment is adopted .
The time $t_{rw}$
necessary  to escape
from the box in the presence  of a random walk is
\begin{equation}
t_{rw} = \frac {L^2}{4 \delta v_{tr}}
\quad.
\end {equation}
The  equation  to be solved is now
\begin{equation}
\frac {\tau_r}{t_{rw}} =1
\quad.
\end {equation}

By imposing $v_{tr}  = \frac{c}{c_f}$,  
where $c_f$ is a  parameter greater than one,
the maximum frequency $\nu_{c,rw}^{max}$ carried by the
electron is
\begin{equation}
\nu_{c,rw}^{max} =\nu_{c}^{max}
\times  (\frac {2}{c_f~NDIM})^2
\quad.
\end{equation}
The relativistic electrons move at velocity 
near to that of the light but the trajectory is not rectilinear.
The parameter $c_f$   averages the irregularities 
and fixes  the transport velocity  once the length  and the 
direction of the step are fixed.
The  presence of the random walk produces  a decrease
in   the  maximum frequency that can be carried by the electron .
For   example,
with   $\nu_{c,rw}^{max}~=4.0~10^{17}~Hz$ ,
and assuming $ c_f$=2  the maximum  value allowed for
NDIM is  1273;
for bigger values of  NDIM  the X-ray emission cannot be
sustained.

\subsubsectionb{4.2.2} {Damping length and gyro-radius}

\label{damping}
The  length travelled by the electron  $L_{rw}$
before  being  damped
is
\begin{equation}
L^2_{rw} = 2 d D t_{rw}
\quad  .
\end{equation}
On replacing  $t_{rw}$  with  it's physical value 
 we obtain the damping length
\begin{equation}
L_{rw} = 110.312 (\frac {L_{pc}}{c_f~NDIM})^{1/2}
        \times ( \frac {1}{H^{3/4} \nu_c ^{1/4}} )~pc
\quad  .
\label{lrwnu}
\end {equation}
The  electron gyro-radius  ,
see 
       Lang (1999)
 , formula~(1.153),  is 
\begin {equation}
\rho  \approx  2\times 10^{9}  E  H^{-1} cm
\quad  ,
\end {equation}
with energy in ergs and $H$ in Gauss . It
can be expressed in pc  by
introducing the critical frequency ( in Hz) ,
( Bohm diffusion)
\begin {equation}
\rho  \approx  2.67 \times 10^{-19} \frac {\nu_c ^{1/2}}{H^{3/2}}
~pc
\label{ro}
\quad  .
\end {equation}
In order to have transport the following inequality
should be satisfied
\begin{equation}
\rho~<~L_{rw}
\quad   .
\label{inequality}
\end {equation}
If equation~(\ref{ro}) is solved   for $\nu$,  the
following is obtained:
\begin {equation}
\nu~<~3.197 \times 10^{27}  H
( \frac {L_{pc} } {c_f~NDIM} )^{2/3} Hz
\quad.
\label{disequalitynu}
\end{equation}
 With  the data from  Sect.~4, for example,
the
inequality is verified:
the  right  hand side of (\ref{disequalitynu})
is  $6.062~10^{20}$~Hz  and the left hand side is
$\nu=4~10^{17} $~Hz.
It is  also possible to test the hypothesis of  
random walk  without appreciable
synchrotron losses;  expressed as an inequality is
\begin{equation}
 L_{rw} > L_{pc} 
\quad.
\end{equation}
Upon inserting the typical  values of the X-region (NDIM=183 
and $\nu=4~10^{17} $~Hz)  formula~(\ref{lrwnu})  gives  
\begin{equation}
 0.62 > L_{pc} 
\quad,
\end{equation}
and as being $L_{pc}$=0.52 , the inequality is verified.
In the  radio region (NDIM=2~$10^7$  
and $\nu=30~10^{6} $~Hz) the application of  formula~(\ref{lrwnu}) 
gives
\begin{equation}
 0.7 > L_{pc} 
\quad,
\end{equation}
and therefore  the inequality is verified.

\subsubsectionb{4.2.3}{Number of collisions on an unbounded lattice}

\label{physics_w}
The random walk in 1D on an unbounded lattice is characterised 
by the length of the step  $\delta$, which in our case is equalised
to the value of the electron's  gyro-radius  $\rho$,
and by the number of steps after which the random motion
is stopped.
The damping time ,$\tau_r$, is given by equation~(\ref{taur}) 
and the time     ,$\tau_{\rho}$ , necessary 
to travel a distance $\delta=\rho$ 
at a velocity  $v_{tr}  = \frac{c}{c_f}$
is  
\begin {equation}
\tau_{\rho} =
{2.66\times 10^{-11}}\,{\frac {\sqrt {H\nu_c}{\it c_f}}{{H}^{2
}}}~sec
\quad  , 
\end {equation}
where $\nu_c$ is expressed in Hz and  H in Gauss.
The number of collisions made by the electron 
before  halving  it's energy is 
\begin{equation}
N = \frac{\tau_r}{\tau_{\rho}} =
{4.699 \times 10^{22}}\,{\frac {1}{\nu_c\,{\it c_f}}}
\quad.
\end{equation}
Table~\ref{data_1D} reports the set of parameters after which the
profile of the concentration in the asymmetric 1D random walk
is similar  to the observed one.
The number of collisions before to damp the relativistic electron
is independent from the magnetic field  but depends 
from  $c_f$ the factor that fixes the transport velocity 
and  $\nu_c$, the frequency of emission.
Is   also possible to compute the mean square 
displacement  $\overline{R^2(t)}$ 
as given 
by equation~(\ref{meansquare})  corresponding to the damping time
$\tau_r$
\begin{equation}
\overline{R^2(\tau_r)}
=
{3.15\times 10^{-15}}\,{\frac {1}{{H}^{3}{\it c_f}}} pc^2
\quad.
\label{r2pc}
\end{equation}
Is interesting  to point out that the mean square displacement  
is independent from  the chosen energy/frequency ( framework 
of the Bohm diffusion) and therefore the observed profiles 
should be independent from the chosen type of astronomy.
This independence from the band of observation 
 can explain Figure~4 of 
        Dyer et~al. (2004)
  where there are minimum
differences between  radio and x profiles in the image of SN1006. 
 \begin{table} 
 \caption[]{The physical data on asymmetric 1D random walk } 
 \label{data_1D} 
 \[ 
 \begin{array}{llc} 
 \hline 
 \hline 
 \noalign{\smallskip} 
 symbol  & meaning & value  \\ 
 \noalign{\smallskip} 
 \hline 
 \noalign{\smallskip}
H_{-4} &  magnetic~field~in~10^{-4}~units  &   1.2              \\ \noalign{\smallskip}
\rho   & electron's~gyro-radius               &   1.24~10^{-4} pc  \\ \noalign{\smallskip}
\nu_c  & critical~frequency~in~X-region      &   4~10^{17} Hz   \\ \noalign{\smallskip}
N      & number~of~collisions                 &   58744         \\ \noalign{\smallskip}
\noalign{\smallskip} 
 \hline 
 \hline 
 \end{array} 
 \] 
 \end {table} 

The assumption
of the mathematical diffusion that sets the concentration equal 
to zero at a given distance from the plane where the diffusing
substance is introduced , is justified by the physical effect here 
described.

\sectionb {5} {Montecarlo diffusion }
\label{MonteCarlo}
A  numerical simulation is now performed    by implementing
the asymmetric random walk of a charged  particle on  a 1D
regular network  consisting of  a lattice with $NDIM$
grid  points embedded  on  a length $side(pc)$;
the conversion between
the unit--length and  physical dimension
of one step is the factor $\frac{side(pc)}{(NDIM -1)}$.
The analysis is divided  into  random walks with absorbing boundaries
and random walks on an infinite or unbounded  lattice.

\subsectionb{5.1}{Absorbing boundaries}

The rules adopted in implementing the 1D asymmetric random walk
with injection in the middle of the grid  are :
\begin {enumerate}
\item    The first of the  NPART  electrons is chosen.
\item    The random  walk of an electron starts in the middle of
         the grid.
         The probabilities of  having  one step 
         are $p_1$ in the
         negative direction  (upstream)
         ,$ p_1 =  \frac{1}{2}- asym \times \frac{1}{2}$,
         and $p_2$ in the positive direction  (downstream)
         , $ p_2 =  \frac{1}{2}+ asym \times \frac{1}{2}$.

\item    When the electron reaches    one of the two 
         absorbing points ,  the motion starts
         another  time from (2) with
         a different diffusing  pattern.

\item    The number of visits is  recorded on ${\mathcal M}$ ,
         a one--dimensional grid.
\item    The random walk terminates when all  the NPART
         electrons are processed.
\item    For the sake  of normalisation the
         one--dimensional visitation/concentration grid
         ${\mathcal M}$ is divided by  NPART.

\end  {enumerate}

These  transition  probabilities  can be
considered the 1D counterpart of the 2D formula
(4.5)  in  
       Ferraro  \&  Zaninetti (2001)
.
The introduction of the asymmetry in the probability
means that it is
possible to simulate diffusion and convection. 

 There is a systematic change of the average particle
position along the
x-direction:
\begin{equation}
\langle dx \rangle =    asym~\delta \quad,
\label{formula1}
\end {equation}
per each time step. 
If the time step is dt=$\frac{\delta}{v_{tr}}$
and the transport velocity
\begin{equation}
v_{tr}  = \frac{c}{c_f}
\quad  ,
\end{equation}
where $c_f$ is a parameter greater than one ,
the convection velocity
$u$ is
\begin{equation}
u= \frac {asym~c} {c_f} \quad.
\label{convection}
\end{equation}

\label{shock}
The density distribution is actually determined by
the ratio $u/D$ rather than $u$ or $D$ separately,
so using a single $u$ and $D$
throughout the system only means that
the ratio is fixed to a constant
value.
Let consider a strong shock moving in the laboratory frame 
with velocity $u_s$.
 The post-shock speed is
\begin{equation}
  u_p = {u_s} \frac{3}{4}
 \quad, 
\label{postshock}
\end {equation}
and the post-shock gas moves in the same 
direction as the shock.
On introducing  the index $u$ and $d$ , denoting
respectively the  upstream   and the  downstream
region,
 the ratio of the  Larmor-radii is
\begin{equation}
\frac{\rho_u}{\rho_d} = \frac{H_d}{H_u} = {4}
\quad,
\end{equation}
, see equation~(9.8b) and  equation~(9.12) in 
       McKee (1987)
.

Let now consider a frame in which the shock is at rest .
The downstream velocity , $u_{ds}$,
is  
\begin{equation}
u_{ds} =  \frac {3}{4} u_s - u_s  =-  \frac{1}{4} u_s
\quad.
\end{equation}
The downstream  velocity is negative and is 1/4 of the shock velocity;
of course if we take an  x-axis  with positive direction toward 
the center of SNR this velocity is positive.

Thus, using a single value for both $u$ (the advection velocity)
and $D$ (the diffusion coefficient) 
may be justified,
but for these values
 the downstream  velocity $\frac{u_s}{4}$
= 650  $km~s^{-1}$
and the
downstream  $\rho_d$  (which   corresponds
to the magnetic field of equipartition
and the largest radiative region)
for $\delta$
should be used.

\begin{figure*}
\begin{center}
\includegraphics[width=12cm,angle=-90]{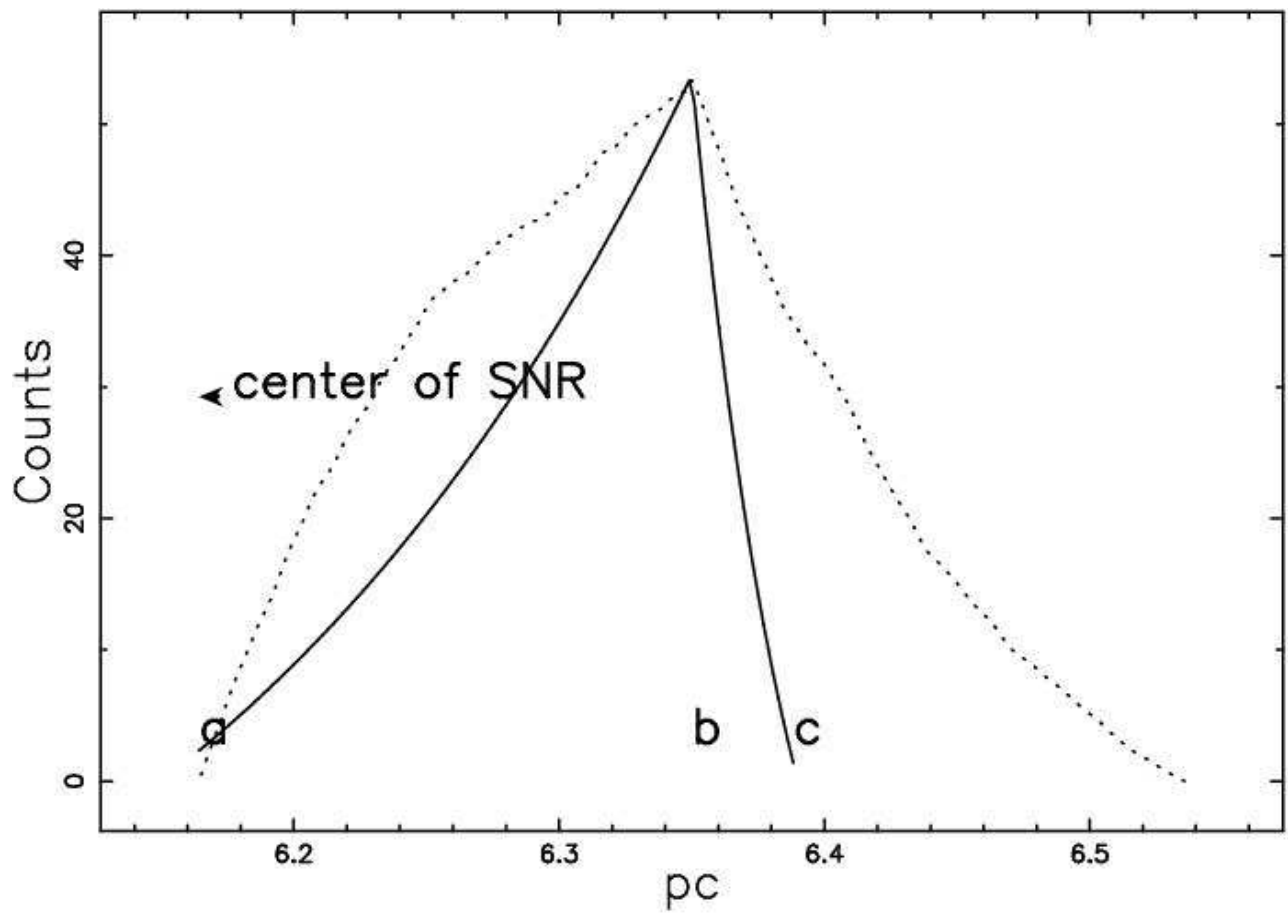}
\end {center}
\caption
{
Observed profile of the  non-thermal X filament in SN 1006 
(full line ) and  concentration of the 1D asymmetric random walk 
(dotted  line), NDIM=183 ,NPART=2000 , $c_f$=6  and
$asym$ =0.013. 
Parameters  as  in Table~1.
}
\label{f07}
    \end{figure*}

The
expected value of {\it asym} can now be
easily deduced from
equation~(\ref{convection})
by inserting  $u$=$\frac{u_s}{4}$~:
\begin{equation}
asym = c_f * 0.0021
\quad   ,
\end{equation}
and Figure~\ref{f07} reports the observed X-profile as well
as the  profile of concentration as given by the 
1D random walk with drift.
A formula for the ratio $\frac{u}{D}$ , where $u$ represents
the drift velocity, can be derived in the X-region
\begin{equation}
\frac {u}{D} = 2 \frac{u}{c} c_f \frac{NDIM}{side} = 1.525 c_f ~pc^{-1}
\label{formulaud}
\quad.
\end{equation}

\subsectionb{5.2}{Unbounded  lattice}

\begin{figure*}
\begin{center}
\includegraphics[width=12cm,angle=-90]{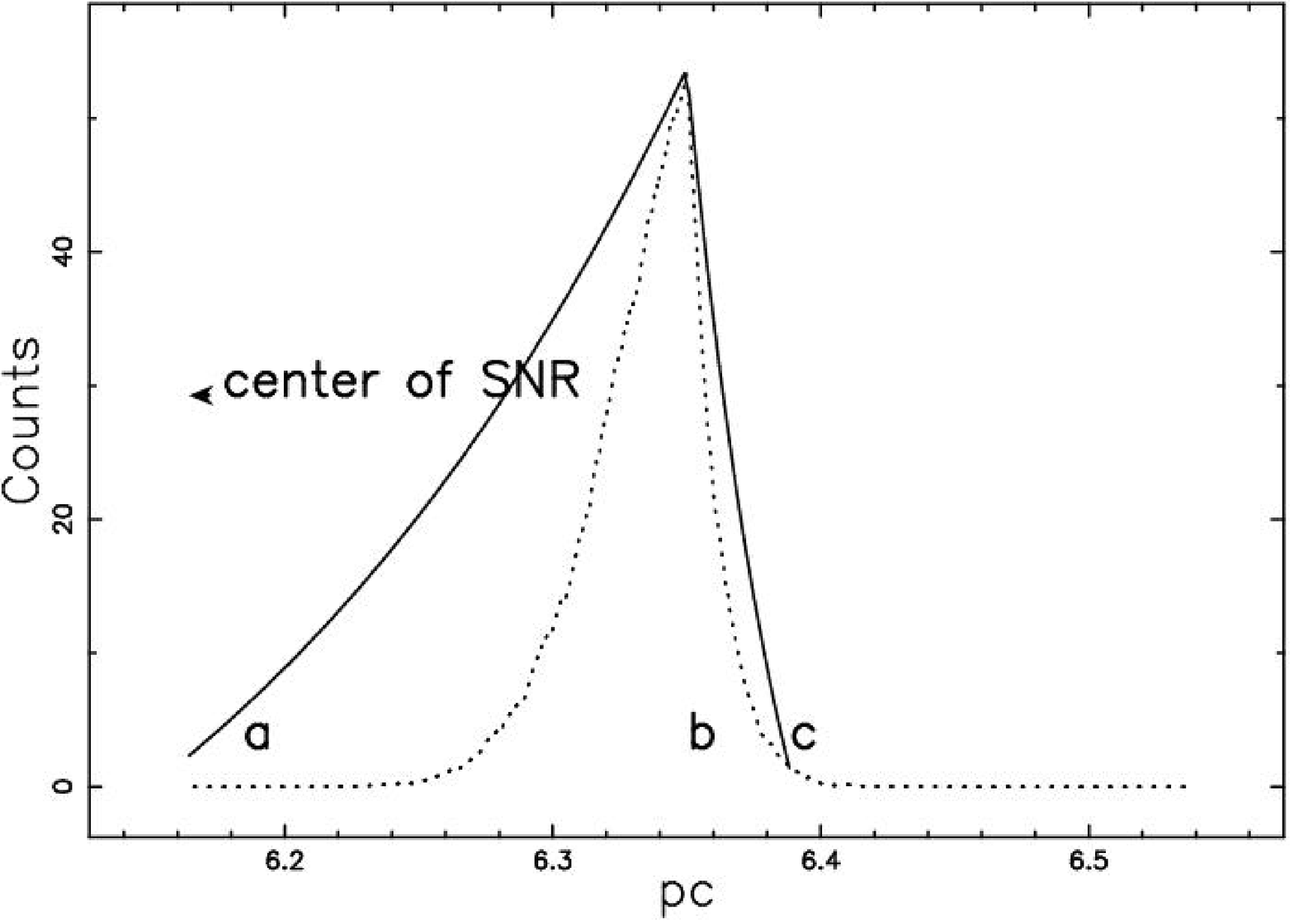}
\end {center}
\caption
{
Observed profile of the  non-thermal X filament in SN 1006 
(full line ) and  1D asymmetric random walk 
(dotted  line), NDIM=2980,NPART=2000 , $c_f$=2, N= 58744 and
$asym$=0.0043. 
Physical parameters as  in Table~2.
}
\label{f08}
    \end{figure*}

The procedure that describes the random walk is the same  
as the previous subsection except for  
the fact that the random walk at item  (3) stops after N steps rather
than  when one of the two limiting boundaries is  reached.
Figure~\ref{f08} reports the observed X-profiles as well
as the  profiles of concentration as given by the 
1D random walk with drift  according to the physics explored
in Sect.~4.2.3.

\sectionb {6} {Theory of the image}

\label{intensity}
\label{sec_x} 
The overall  behaviour
of SN1006 in the
X--region , see for example  
       Dyer et~al. (2004)
,
presents a
spherical symmetry  with a ring enhancement
and a central
depression. When these effects are visualised through a cut in the
flux crossing the center of the image, a characteristic "U" profile
is obtained.
These "U" profiles can be explained through
cuts of the intensity given by the 3D mathematical diffusion
while  the agreement with the theoretical predictions excludes 
the presence of absorption. 

The asymmetry in the observed flux can be simulated
through the map 
of the field of  velocity~.

\subsectionb {6.1}{The mathematical image}

\begin{figure*}
\begin{center}
\includegraphics[width=12cm,angle=-90]{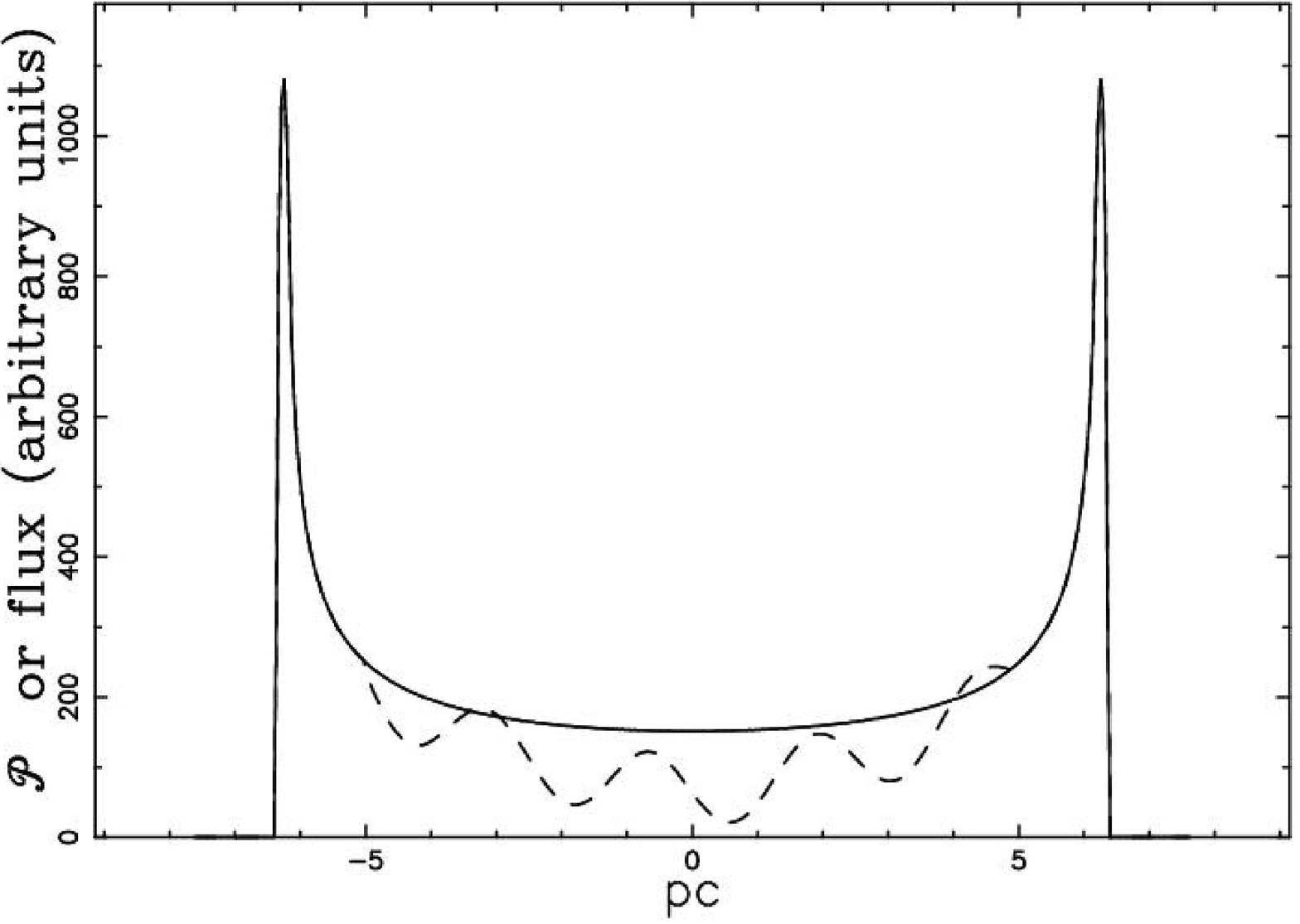}
\end {center}
\caption
{
 Cut of the mathematical  intensity ${\it I}$
 crossing the center    (full line  )
 and X-ray data (dashed line)
 extracted from
       Dyer et~al. (2004)
.
 Parameters as  in Table~1, great box.
}
\label{f09}
    \end{figure*}

The observed flux of the overall morphology of the SNR is
presented in the form of "U" profile , see dashed line in
Figure~\ref{f09}. 
The theoretical intensity can be found by 
solving the integral equation ( \ref{transport}) along the x direction
 , see Figure \ref{f02} where three lines of sight  are
 reported. 

The concentrations to be used
are formulae (\ref{cab}) and  (\ref{cbc})  once
$r=\sqrt{x^2+y^2}$ is imposed ;
these two concentrations are  inserted in
formula~(\ref{eqn_transfer})  which  represents the transfer equation.
 The     geometry of the phenomena fixes
 three different zones ($0-a,a-b,b-c$) in the variable $y$;
 the first piece , $I^I(y)$ , is
\begin{eqnarray}
 I^I(y)= \int _{\sqrt{a^2-y^2}} ^{\sqrt{b^2-y^2}} 2 C_{ab}dx
 + \int _{\sqrt{b^2-y^2}} ^{\sqrt{c^2-y^2}}  2C_{bc}dx \nonumber\\ 
~= 
 2\,{\frac {b{\it Cm}\, \left( -\sqrt {{a}^{2}-{y}^{2}}+a\ln  \left(
\sqrt {{a}^{2}-{y}^{2}}+a \right) +\sqrt {{b}^{2}-{y}^{2}}-a\ln
 \left( \sqrt {{b}^{2}-{y}^{2}}+b \right)  \right) }{b-a}}\nonumber\\ 
~-2\,{\frac {
b{\it Cm}\, \left( c\ln  \left( \sqrt {{b}^{2}-{y}^{2}}+b \right)
- \sqrt {{b}^{2}-{y}^{2}}-c\ln  \left( \sqrt {{c}^{2}-{y}^{2}}+c
 \right) +\sqrt {{c}^{2}-{y}^{2}} \right) }{c-b}} \nonumber\\ 
~ 0 \leq y < a \quad.
\label{I_1}
\end{eqnarray}

The second piece , $I^{II}(y)$ , is
 \begin{eqnarray}
 I^{II}(y)=  \int _0 ^{\sqrt{b^2-y^2}} 2 C_{ab}dx
 + \int _{\sqrt{b^2-y^2}} ^{\sqrt{c^2-y^2}}  2C_{bc}dx \nonumber\\ 
~= 
-{\frac {b{\it Cm}\, \left( -a\ln  \left( {y}^{2} \right)
-2\,\sqrt {{b}^{2}-{y}^{2}}+2\,a\ln  \left( \sqrt
{{b}^{2}-{y}^{2}}+b \right)
 \right) }{b-a}} \nonumber\\ 
~-2\,{\frac {b{\it Cm}\, \left( c\ln  \left( \sqrt {{b}
^{2}-{y}^{2}}+b \right) -\sqrt {{b}^{2}-{y}^{2}}-c\ln  \left(
\sqrt {{c}^{2}-{y}^{2}}+c \right) +\sqrt {{c}^{2}-{y}^{2}}
\right) }{c-b}} \nonumber\\ 
  a \leq y < b  \quad. 
\label{I_2}
\end{eqnarray}
The third  piece , $I^{III}(y)$ , is
 \begin{eqnarray}
 I^{III}(y)= \int_0 ^{\sqrt{c^2-y^2}}  2C_{bc}dx \nonumber\\ 
~= 
{\frac {b{\it Cm}\, \left( -c\ln  \left( {y}^{2} \right) +2\,c\ln
 \left( \sqrt {{c}^{2}-{y}^{2}}+c \right) -2\,\sqrt {{c}^{2}-{y}^{2}}
 \right) }{c-b}}
\nonumber\\ 
 b \leq y < c  \quad. 
\label{I_3}
\end{eqnarray}

\begin{figure*}
\begin{center}
\includegraphics[width=12cm,angle=-90]{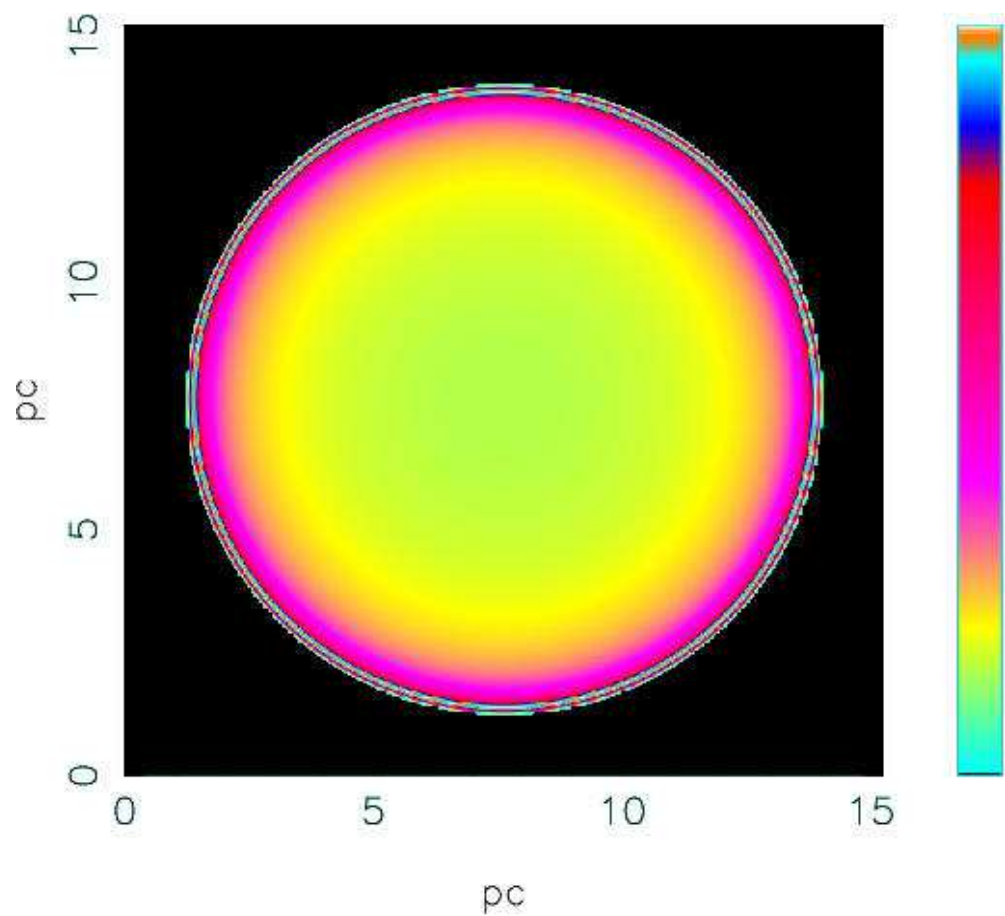}
\end {center}
\caption
{
Contour map  of  ${\it I}$
particularized to simulate the X-emissivity of SN~1006.
Parameters as  in Table~1, great box.
}
\label{f10}
    \end{figure*}

The profile  of ${\it I}$   made by the three 
pieces (~\ref{I_1}), (~\ref{I_2}) and  (~\ref{I_3}),
can then be plotted as a function of the distance from
the center , see~Figure~\ref{f09},
or like a   contours , 
see~Figure~\ref{f10}.

The position of  the minimum
of ${\it I}$  is at $y=0$ and the position of the maximum
is situated in the region $a \leq y < b $,
and more precisely at:
\begin{equation}
y={\frac {\sqrt {- \left( b-2\,a+c \right) a \left( ab-2\,bc+ac
\right) }}{b-2\,a+c}}
\quad. 
\label{maximum} 
\end{equation}

The ratio between the theoretical intensity at maximum , $I_{max}$ ,
as given by formula~(\ref{maximum}) and at minimum ($y=0$) 
is given by 

\begin{equation}
\frac {I_{max}} {I(y=0)} = \frac{Numerator}{Denominator}
\quad,
\label{ratioteor}
\end {equation}
where 
\begin{eqnarray}
 Numerator= \nonumber\\ 
 -a\ln  \left( -{\frac {\left( ac+ab-2\,cb \right) a}{b+c-2\,a}}
 \right) c+a\ln  \left( -{\frac {\left( ac+ab-2\,cb \right) a}{b+c-2
\,a}} \right) b-2\,\sqrt {{\frac {\left( c+b \right)  \left( b-a
 \right) ^{2}}{b+c-2\,a}}}c \nonumber\\ 
 -2\,a\ln  \left( \sqrt {{\frac {\left( c+b
 \right)  \left( b-a \right) ^{2}}{b+c-2\,a}}}+b \right) b+
\nonumber  \\
2\,c\ln
 \left( \sqrt {{\frac {\left( c+b \right)  \left( b-a \right) ^{2}}{b
+c-2\,a}}}+b \right) b+2\,\sqrt {{\frac {\left( c+b \right)
\left( b -a \right) ^{2}}{b+c-2\,a}}}a
\nonumber \\
 -2\,c\ln  \left( \sqrt
{{\frac {\left( a -c \right) ^{2} \left( c+b \right)
}{b+c-2\,a}}}+c \right) b+ \nonumber \\
2\,c\ln
 \left( \sqrt {{\frac {\left( a-c \right) ^{2} \left( c+b \right) }{b
+c-2\,a}}}+c \right) a+2\,\sqrt {{\frac {\left( a-c \right) ^{2}
 \left( c+b \right) }{b+c-2\,a}}}b-
 \nonumber \\
 2\,\sqrt {{\frac {\left( a-c
 \right) ^{2} \left( c+b \right) }{b+c-2\,a}}}a
 \quad,
\end{eqnarray}
and 
\begin{eqnarray}
 Denominator= \nonumber\\ 
-2\,ac\ln  \left( a \right) +2\,ba\ln  \left( a \right) -2\,ba\ln 
 \left( b \right) +2\,bc\ln  \left( b \right)  
 -2\,bc\ln  \left( c
 \right) +2\,ac\ln  \left( c \right)
\quad.
\end{eqnarray}
In order to make the model more realistic
the maximum  value of ${\it I}$  is
normalised  in order to have  the same
value as  the flux in arbitrary units    (i.e. 1080)
in  Figure~4
from  
       Dyer et~al. (2004)
.

From  the simulation  reported  in
Figure~\ref{f09}  we obtain
\begin {equation}
\frac {{\it I}_{ring}} {{\it I}_{center}} = 7.14
\quad,
\label{ratiovalue}
\end {equation}
where  ${\it I}_{ring}$ represents the  maximum value
of  ${\it I}$ ( on the ring)  and
${\it I}_{center}$ the value  at the center.
The X-observations
with   ASCA
of SN~1006 , see Figure~4 by  
       Dyer et~al. (2004)
  ,
give
\begin {eqnarray}
\label{values}
\frac {flux_{ring}} {flux_{center}} = 1080/200 = 5.4 &~minimum~value
\nonumber \nonumber\\
~\nonumber\\
\frac {flux_{ring}} {flux_{center}} = 1080/100 = 10.8&~maximum~value
\quad   ,             \nonumber
\end {eqnarray}
here $flux_{ring}$ and  $flux_{center}$ are
the two fluxes  in arbitrary units
( 200 is the maximum value of the
 fluxes in the central region
 and  100 is an   approximated average value  of the  fluxes
 in the central region~).

The    mathematical diffusion 
 here adopted
predicts   a value  given
$\frac {flux_{ring}} {flux_{center}}=7.14$,
which  is comprised
in the reasonable interval  of observations   given
in equation~(\ref{values}).

It is   also  interesting to note that
the  CAP and CENTER region models
( subsets of the SRESC model)
adopted in   Figure~4 by  
       Dyer et~al. (2004)
  predict ,roughly speaking, a  central
flux that  is
\begin {equation}
\frac {flux_{ring}} {flux_{center}} = 1080/400 = 2.7~.
\end{equation}
This  value is three times bigger than  that observed  or,
in other words,  over-predicts the non-thermal emission
at the center of the SNR.

\begin{figure*}
\begin{center}
\includegraphics[width=12cm,angle=-90]{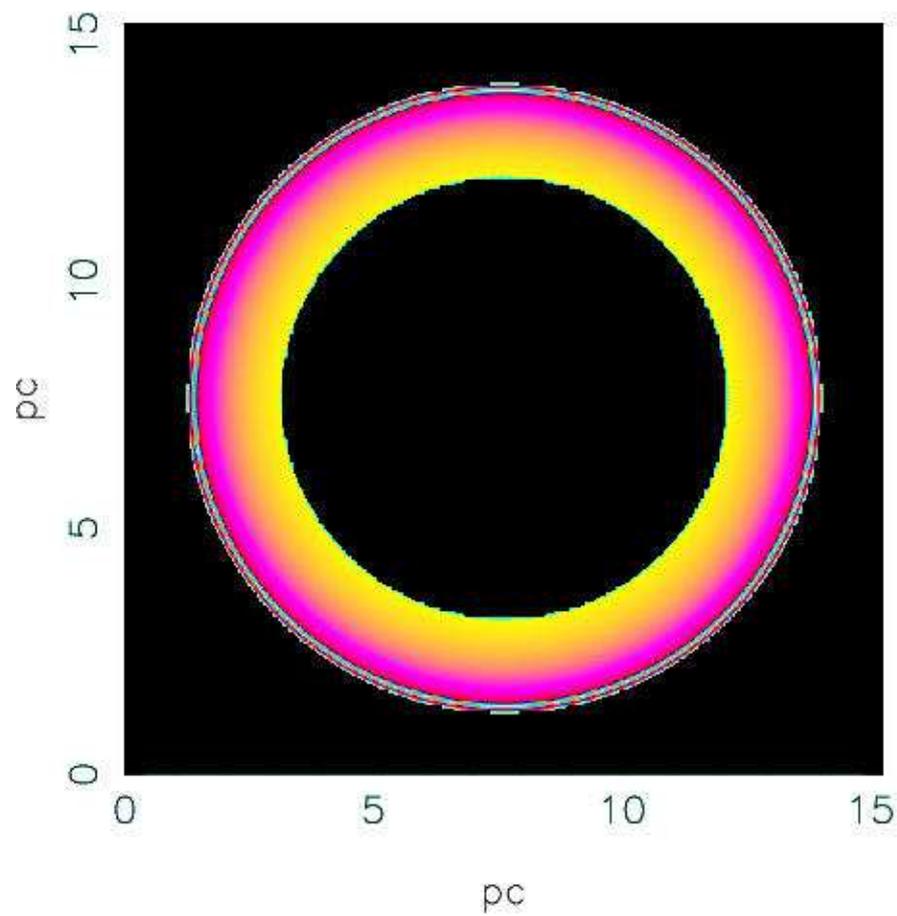}
\end {center}
\caption
{
 The same  as  Figure~\ref{f10}  but with
 $flux_{tr}= flux_{max}/5$
}
\label{f11}
    \end{figure*}

\begin{figure*}
\begin{center}
\includegraphics[width=12cm,angle=-90]{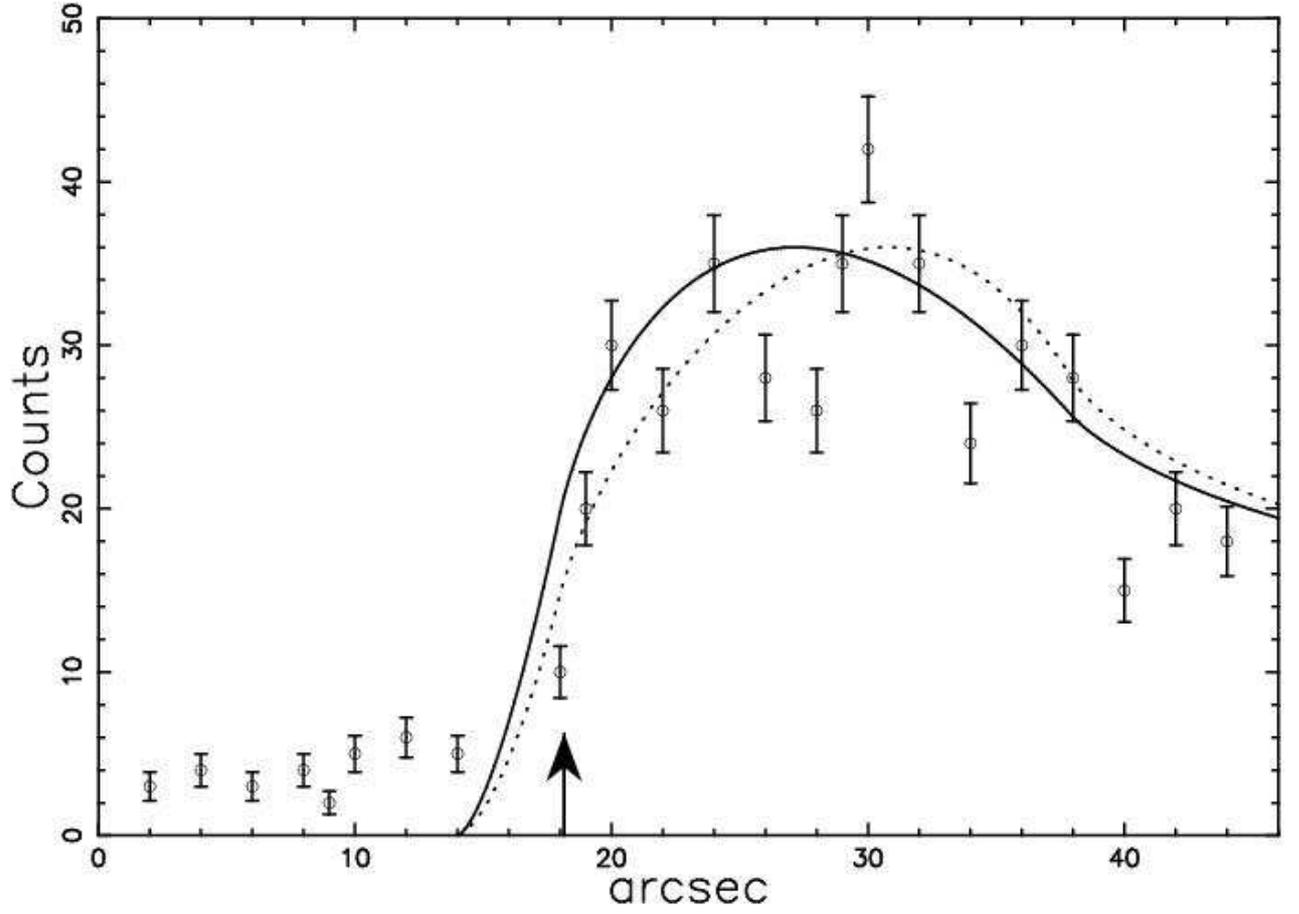}
\end {center}
\caption
{
Cut of the mathematical  intensity ${\it I}$
, full line, that covers a spatial zoom of Figure~\ref{f09}
but with a different scaling.
The radius of the expanding SNR , {\it b} , 
is marked with an arrow.
The experimental data are extracted by the author from 
Figure 4~(panel 1a) in
       Bamba  et~al. (2003)
 .
The dotted line  represents the asymmetric case with  
$\frac {u}{D}=12.2~pc^{-1}$.
}
\label{f12}
    \end{figure*}

The effect of  insertion
of a threshold flux  , $flux_{tr}$, given by the
observational techniques , is now analysed.
The threshold flux  can be parametrised  to  $flux_{max}$,
the maximum  value  of flux characterising
the ring: a typical  image with  a hole  is visible
in  Figure~\ref{f11} when  $flux_{tr}= flux_{max}/5$.
The X-ray emission from SNR  can also be presented through 
cuts in particular  regions, 
see for example Figure~4  in 
       Bamba  et~al. (2003)
, where the x-axis is reported 
in arc-sec and X-intensity in counts.
We also particularize our results in those units , see 
Figure~\ref{f12}
where is possible to visualise the position of the maximum 
of the intensity that is situated between {\it a} and  {\it b}.
The  1D solutions of the concentration in presence of drift 
represented by equation~(\ref{cab_drift}) and (\ref{cbc_drift})  cannot  be integrated  
along  the line of sight in terms of elementary 
functions and therefore a numerical integration is performed,
see  dotted line in Figure~\ref{f12}.

\subsectionb{6.2}{The case of absorption}

\begin{figure*}
\begin{center}
\includegraphics[width=12cm,angle=-90]{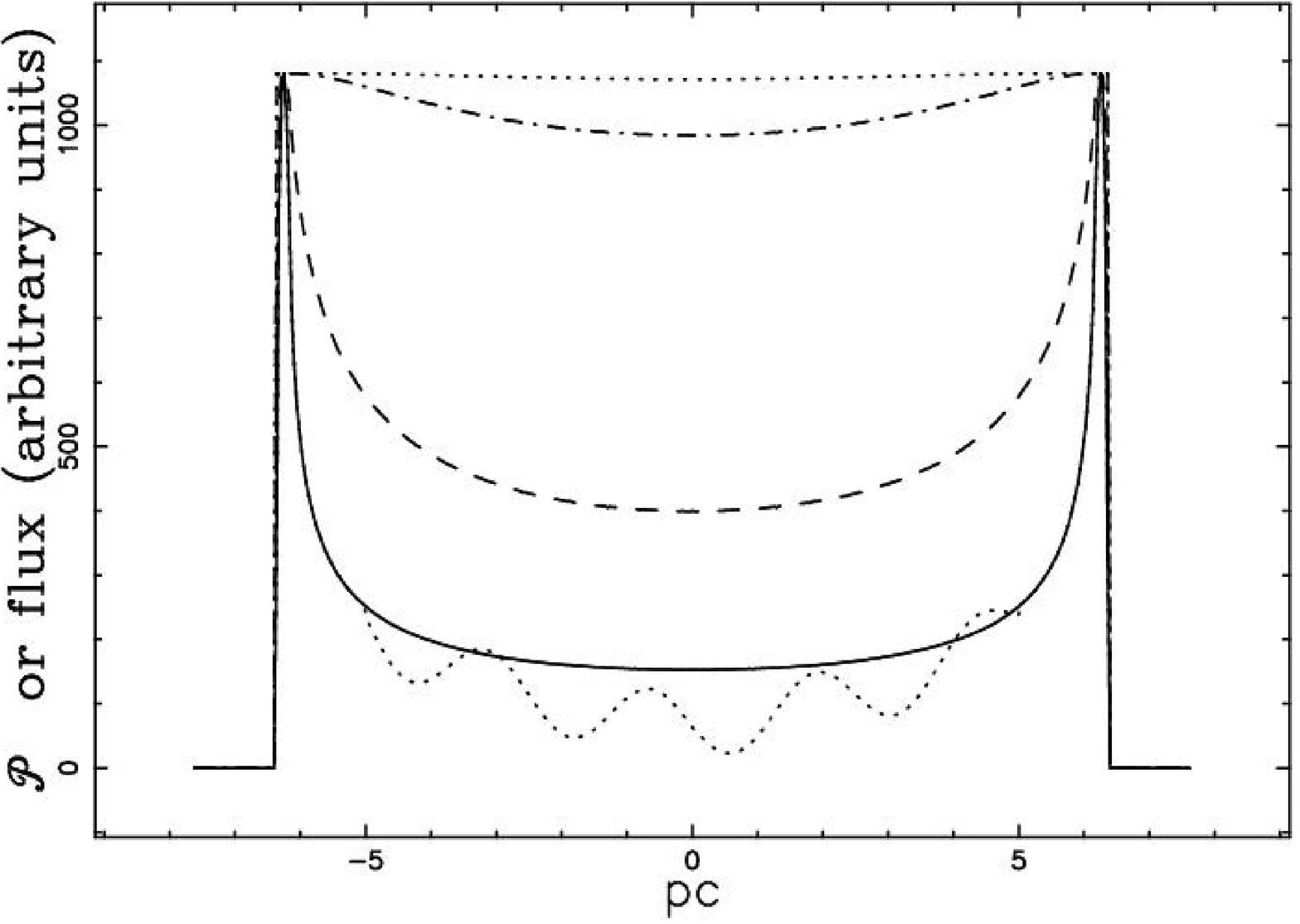}
\end {center}
\caption
{
 Cuts of the projection grid ${\it I}$
 crossing the center:
 X-ray data extracted from
        Dyer et~al. (2004)
    (dashed thin line ),
 $K_a$ = $10^{-2}$          (full line        ),
 $K_a$ = 1.8                (dashed           ),
 $K_a$ = 10                 (dot-dash-dot-dash),
 $K_a$ = 20                 (dotted).
 Parameters as in
 Figure~\ref{f10}, great box.
 Transition from optically thin to  thick layer.
}
\label{f13}
    \end{figure*}

\label{sec_absorption}  
The effect of  absorption is easily  implemented 
by adopting  formula~(\ref{transition})  and fixing 
the value of $K_a$~for which the results of the thin
layer approximation are reproduced.
The value of  $K_a$ is then gradually increased in order to
see how the effect of the absorption  modifies the "U" profiles
in the cut of the intensity previously reported.
This numerical experiment is visible 
in Figure~\ref{f13} in which the transition
from a  "U" profile corresponding to the thin-thick case   is analysed.

Due to the fact that we do not know if  absorption
is present or not,  all the cuts are normalised
in order to have the same maximum value (1080).
This transition explains  the anomalous value
of  $\frac {flux_{ring}} {flux_{center}}$= 2.7
of the CAP and CENTER codes  as due to a high
value adopted  in  the absorption's coefficient.
In  Figure~\ref{f13}  the third (dashed) line
reproduces this anomalous profile.
We conclude this paragraph underlying the fact that 
the absorption is not relevant
in  the X-ray
frequencies .

\subsectionb {6.3}{An unbounded lattice and derivation of the magnetic field}

\begin{figure*}
\begin{center}
\includegraphics[width=12cm,angle=-90]{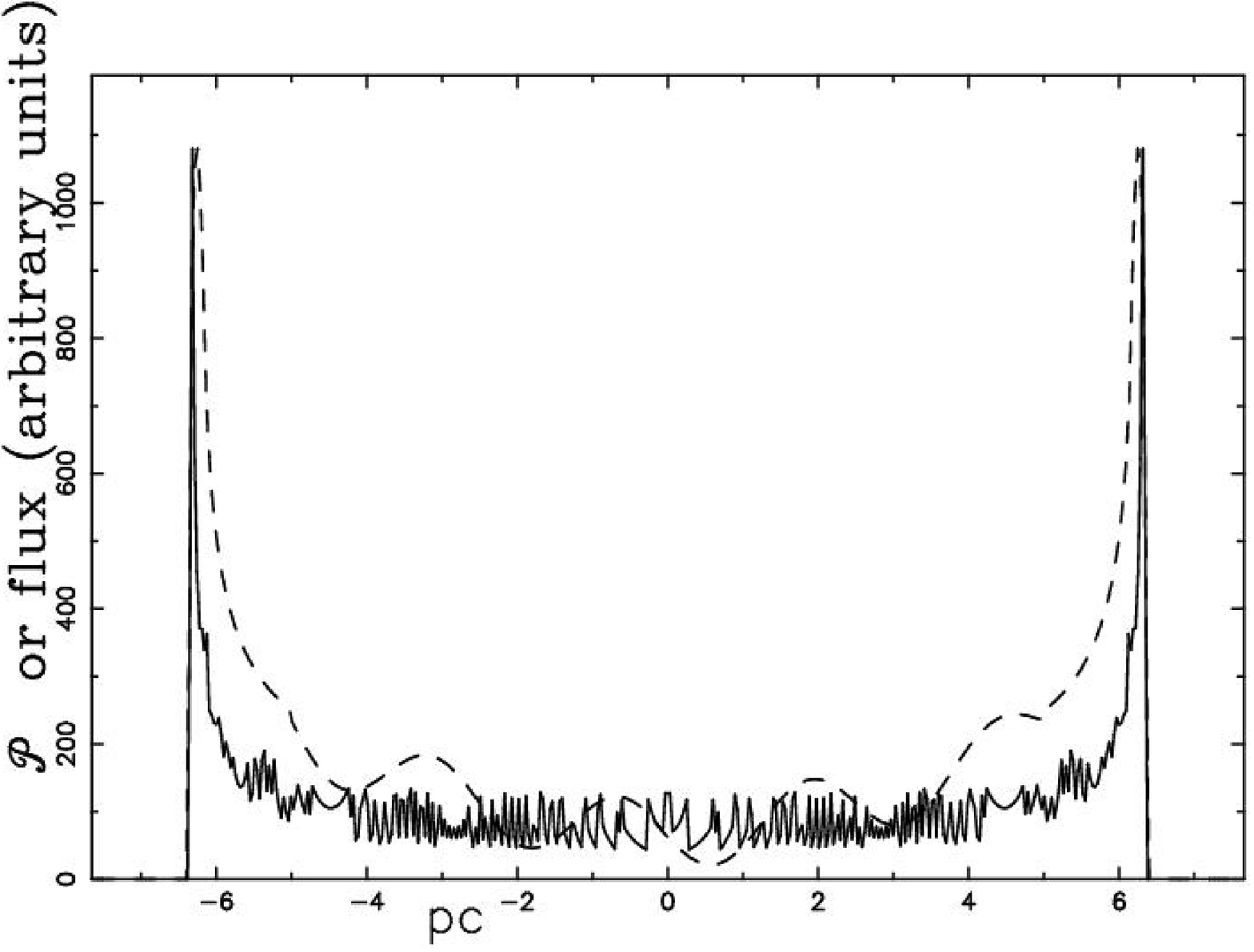}
\end {center}
\caption
{
 Cut of the mathematical  intensity ${\it I}$
 crossing the center    (full line   )
 and X-ray data         (dashed line )
 extracted from   
        Dyer et~al. (2004)
.
 Parameters as  in Table~2, $c_f$=2, but $H_4$=1.2 .
  The simulation returns 
  $W_u^{rw} =1.17~10^{-2}~pc$ and $W_d^{rw}=3.8~10^{-2}~pc$.
}
\label{f14}
    \end{figure*}

\label{unbounded}
The case of an unbounded lattice in the presence of asymmetry 
and the number of collisions fixed by the damping length produces 
profiles in the upstream and downstream region that are reported in 
Figure~\ref{f08}.
These profiles can be parameterised through 
exponentials  of the type of those reported 
in equation~(\ref{exponential}) characterised 
by  $W_u^{rw}$ and   $W_d^{rw}$.
This analytical expression of the concentration of relativistic 
electrons can be the base of the numerical  integral 
along the line of  sight  
with the algorithm outlined in equation~(\ref{thin}).
The typical behaviour of the intensity as a function 
of the distance from the center 
is reported in Figure~\ref{f14}~.

In this case the simulation predicts 
a value  given by 
$\frac {flux_{ring}} {flux_{center}}=9.61$,
which  is comprised
in the  interval  of observations   given
in equation~(\ref{values}).
The magnetic field adopted in this paragraph is independent 
from equipartition arguments and therefore the algorithm
here outlined can be considered a new way to deduce the magnetic field.
 
\subsectionb {6.4}{Comparison with other methods}

In a study on the 
amplification of the magnetic field in SNR, 
the observed/theoretical analysis
    on the radial  profile  of RCW86 in the X-ray band
    ,see Figure 7 in 
           V\"{o}lk et~al. (2005)
     ,
    gives  $\frac {flux_{ring}} {flux_{center}} \approx $4.5.

In the   appendix A of a  paper on X-ray synchrotron emission 
from SNR , 
       Ballet (2006)	
,  
is possible to find a formula  (analogous to our formula
(\ref{ratioteor})) 
that gives the  ratio between the brightness at the center of the
sphere and at maximum ; his Figure A1 
gives  $\frac {flux_{ring}} {flux_{center}} \approx $3~.

\subsectionb {6.5}   {The gamma emission}

Up to now we have focused the attention on the X-ray emission
from SNR ; the gamma ray emission ( $30 MeV < E < 30 GeV$) 
is now analysed  from two different points of view.

\subsubsectionb{6.5.1}{The gamma emission from relativistic electrons}

Taking the lower limit of the gamma emission 
, E=30 Mev  , the corresponding critical  frequency is
$\nu_c~=7.44~10^{21}~Hz$ and the relativistic gyro-radius 
with a magnetic field of  $H_{-4}$=1.2 , 
see Sect.~4.2.3,  $\rho~=0.017~pc$.
Due to the short time of synchrotron losses , N=1  collisions,
the scale widths in the upstream and downstream direction are equal,
$W_u=W_d=0.037~pc$.
In this case the simulation predicts
a value  given by
$\frac {flux_{ring}} {flux_{center}}=7.54$.

\subsubsectionb{6.5.2}{The gamma emission from cosmic rays}
 
The cosmic rays  with energy range  
$0.1~GeV < E < 4 10^5 GeV$, 
see 
       Hillas (2005)
and
       Wolfendale (2003)
 can produce gamma  emission 
from the interaction  with the target material.
Once the CR  energy   is expressed in $10^{15}$eV 
units ( $E_{15}$) , the magnetic field in $10^{-6}$ Gauss ($H_{-6}$)
the relativistic ion's gyro-radius is :
\begin {equation}
\rho_Z = 1.08   \frac {E_{15}}  {H_{-6} Z  } pc 
\label{rho}
\quad,
\end   {equation}
where $Z$ is  the atomic number.
The maximum energies that allow  to apply the random walk 
of CR in shells of SNR are computed.
The magnetic field of equipartition $H_6=15$ ,
the minimum width of the X-profile , $W_u=0.04~pc$ 
and the proton , $Z=1$, 
are selected.
In order to sustain the random walk the following inequality 
should be verified 
\begin{equation}
\rho_Z~<~W_u
\quad,
\end{equation}
that transformed in CR energy 
is 
\begin{equation}
E_{15}~<~0.55 \quad or \quad E~<~5.5~10^5~GeV 
\quad.
\end{equation}
This upper limit is near to the maximum  energy of CR that 
produces gamma radiation , $4~10^5~Gev$.
The various physical processes that allows to produce  
gamma radiation from CR are now summarised :

\begin{itemize}
\item  Acceleration of CR at the shock discontinuity 
       up to $E$ $\approx 5\, 10^5 GeV$,
\item  Diffusion of CR with step's length lower than
       $W_u$, 
\item  Absorption at  $W_u$ and $W_d$ due to a change 
       in the magnetic field from $H_4$=0.15 in the shell
       to $H_4$=0.01  outside the shell,
\item  Production of gamma radiation in a way proportional
       to the concentration of CR. 
\end {itemize}

\subsectionb{6.6} {Asymmetric SNR }

\label{asymmetric}
The theory of an asymmetric SNR  was developed in 
Sect.~4.1 of 
       Zaninetti (2000)
in which an expansion surface
as a function of a non-homogeneous  ISM was computed:
in the same   paper Figure~8  models  SN1006.
The diffusing algorithm  adopted here is the 3D random walk 
from many injection points ( in the following IP)

\begin {enumerate}
\item    The first IP is chosen 
\item    The first of the  NPART  electrons is chosen.
\item    The random  walk of an electron starts where the
         selected IP is situated. 
         The electron moves in one of the six possible directions.

\item    After N steps the process restarts from (2)

\item    The number of visits is  recorded on ${\mathcal M^3}$ ,
         a three--dimensional grid.
\item    The random walk terminates when all  the NPART
         electrons are processed.
\item    The process restarts from (1) selecting another IP 
\item    For the sake  of normalisation the
         one--dimensional visitation/concentration grid
         ${\mathcal M^3}$ is divided by  NPART.
\end  {enumerate}

The {\it IP}\/  are randomly  selected in 
space, and  the   radius is computed 
by using the method  of  bilinear interpolation on the four 
grid points that surround the selected latitude and longitude,
( 
       Press et~al. (1992)
).
The  radius will be the selected value  
+ {\it R}/24  in order  to generate the  IP where 
the action of the shock is maximum. 

\begin{figure*}
\begin{center}
\includegraphics[width=12cm,angle=-90]{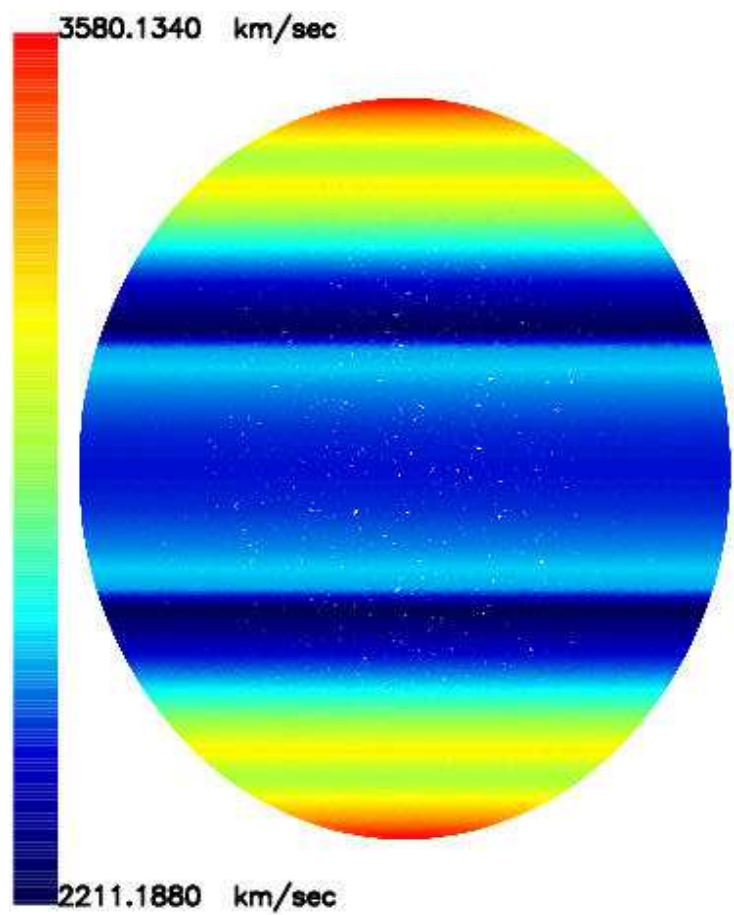}
\end {center}
\caption 
{
  Map of the expansion velocity  
  relative to the simulation of SNR1006 
  when 190000 random points are selected on the surface.
  The   physical parameters are the  same as in 
  Figure~8 of 
         Zaninetti (2000)
  ~.
}
\label{f15} 
    \end{figure*}

Our model gives radial velocities , $V_{theo}$ , 
2211~$km~s^{-1}$ $\leq~V_{theo}~\leq$ 3580~$km~s^{-1}$
and the map of the expansion velocity 
is reported in Figure~\ref{f15}
from which it is possible to visualise the differences 
in the expansion velocities among the various regions
as well as the overall elliptical shape.

Before  continuing  
we should recall that 
in  the presence of discrete time steps on a 3D lattice the average
square radius ,$\langle R^2(N  )\rangle$,  after N steps
(see  
       Gould  \& Tobochnik (1988)
, equation~(12.5~))  is
\begin {equation}
\langle R^2(N  )\rangle  \sim  6 DN \quad,
\label{rquadro}
\end   {equation}
from which the diffusion coefficient , $D$ , is derived
\begin {equation}
D= \frac {\langle R^2(N  )\rangle} {6 N} \quad.
\label{r2N}
\end   {equation}
The two boundaries in which the random walk is  taking 
place are now represented by two irregular surfaces.
It is  possible to simulate them by stopping the random walk  
after a number of iterations $N$ given by  
\begin {equation}
N = NINT (\frac {\overline R_{pc}}{24}\frac  {1}{\delta})^2 
\quad, 
\end {equation}
where $\overline R_{pc}$ represents the averaged radius in pc .
These are the iterations  after which according, 
to formula~(\ref{r2N}),  the walkers reach the boundaries at 
a radial distance given by  $\frac {\overline R_{pc}}{24}$
from the place of injection; 
in other words we are working
on an unbounded  lattice . 
The influence of velocity  
on the  flux  F of radiation  can be inferred from 
the suspected  dependence  when  non-thermal emission
is considered, see equation~(9.29)  in
       McKee (1987)
,  
\begin{equation}
 F= \chi_t \frac {1}{4} \mu_H n_0 v_s^3  
\quad, 
\end {equation}
where $\chi_T$ represents  the efficiency of 
conversion of the unitarian flux of kinetic energy,
$\mu_H$ the mass of the hydrogen nucleus,
$n_0$ the  particles/$cm^3$ and 
$v_s$ the velocity of the shock.
 
\begin{figure*}
\begin{center}
\includegraphics[width=12cm,angle=-90]{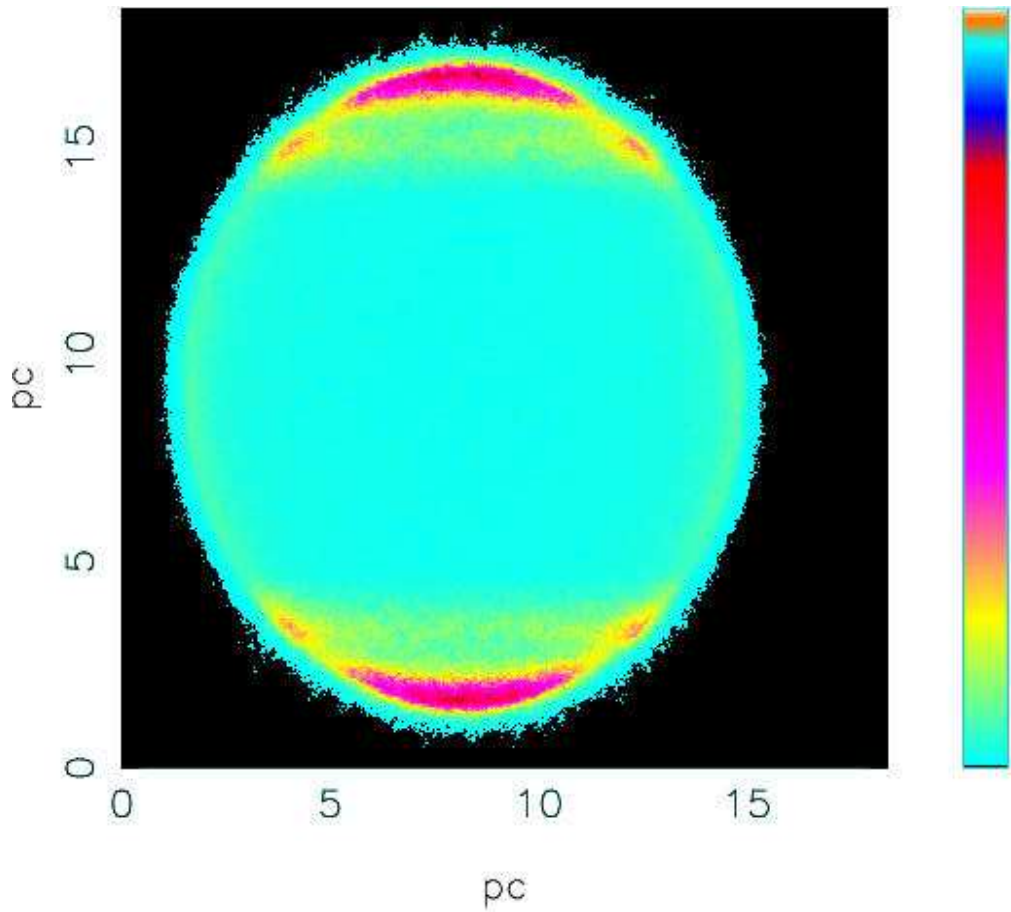}
\end {center}
\caption
{
 Contour  of the intensity {\it I} in the X-rays.  
 The parameters are 
 $ side_{SNR}$=18.37 pc ,
 $\delta=6.12~10^{-3}$~pc,
 $\rho  =2.8~10^{-3}$~pc,
 NDIM=3001,
 $IP=190000^2$ , 
 {\it NPART= 100}, 
 $NTIMES_{max}$=18,
 great box.
Optically thin layer.
}
\label{f16}
    \end{figure*}

\begin{figure*}
\begin{center}
\includegraphics[width=12cm,angle=-90]{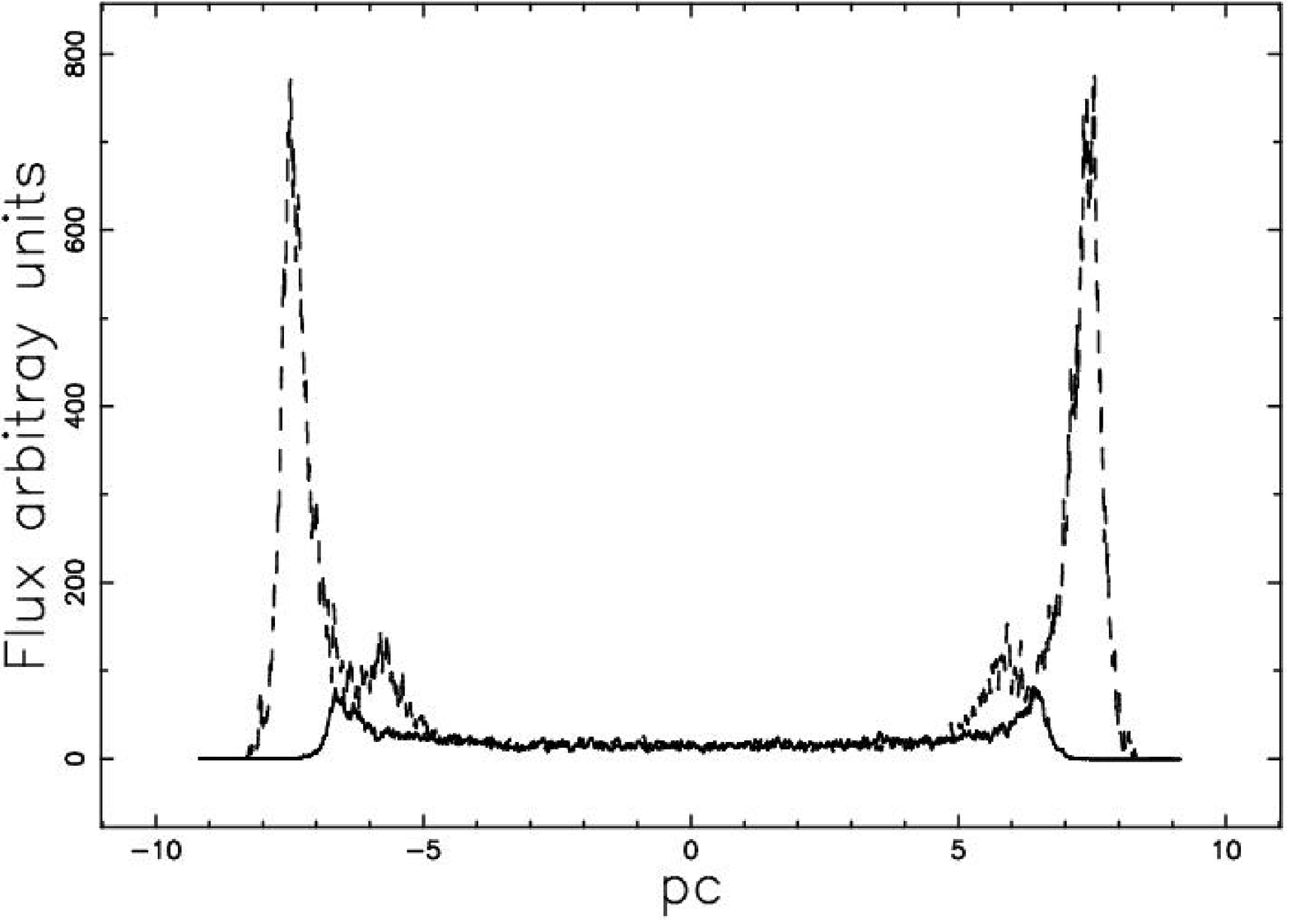}
\end {center}
\caption
{
 Two cut  along  perpendicular lines  of
 {\it I } 
 reported  in Figure~\ref{f16}~.
Optically thin layer.
}
\label{f17}
    \end{figure*}

Assuming that the flux reversed in the 
non--thermal emission follows a similar law
through the parameter $\chi_X$ ( the efficiency in the X--region)
the effect of velocity is simulated through the following
algorithm.
Once  the IP are  spatially  generated, the number of times 
NTIMES 
over which to repeat the cycle is given
by
\begin{equation}
NTIMES = 1 +NTIMES_{MAX} *
( \frac {v - v_{min}}{v_{max}-v_{min}})^3  
\quad, 
\end {equation}
where  $NTIMES_{MAX}$
 is the maximum  of the allowed values
of  NTIMES minus 1,   and v is  the velocity associated
to each IP.
The asymmetric contour map obtained when
the spatial step is $\approx$ 2 * gyro--radius  
is  reported  in Figure~\ref{f16}
and  the cut along two perpendicular lines
of the projection grid in 
Figure~\ref{f17}.
%

\begin{figure*}
\begin{center}
\includegraphics[width=12cm,angle=-90]{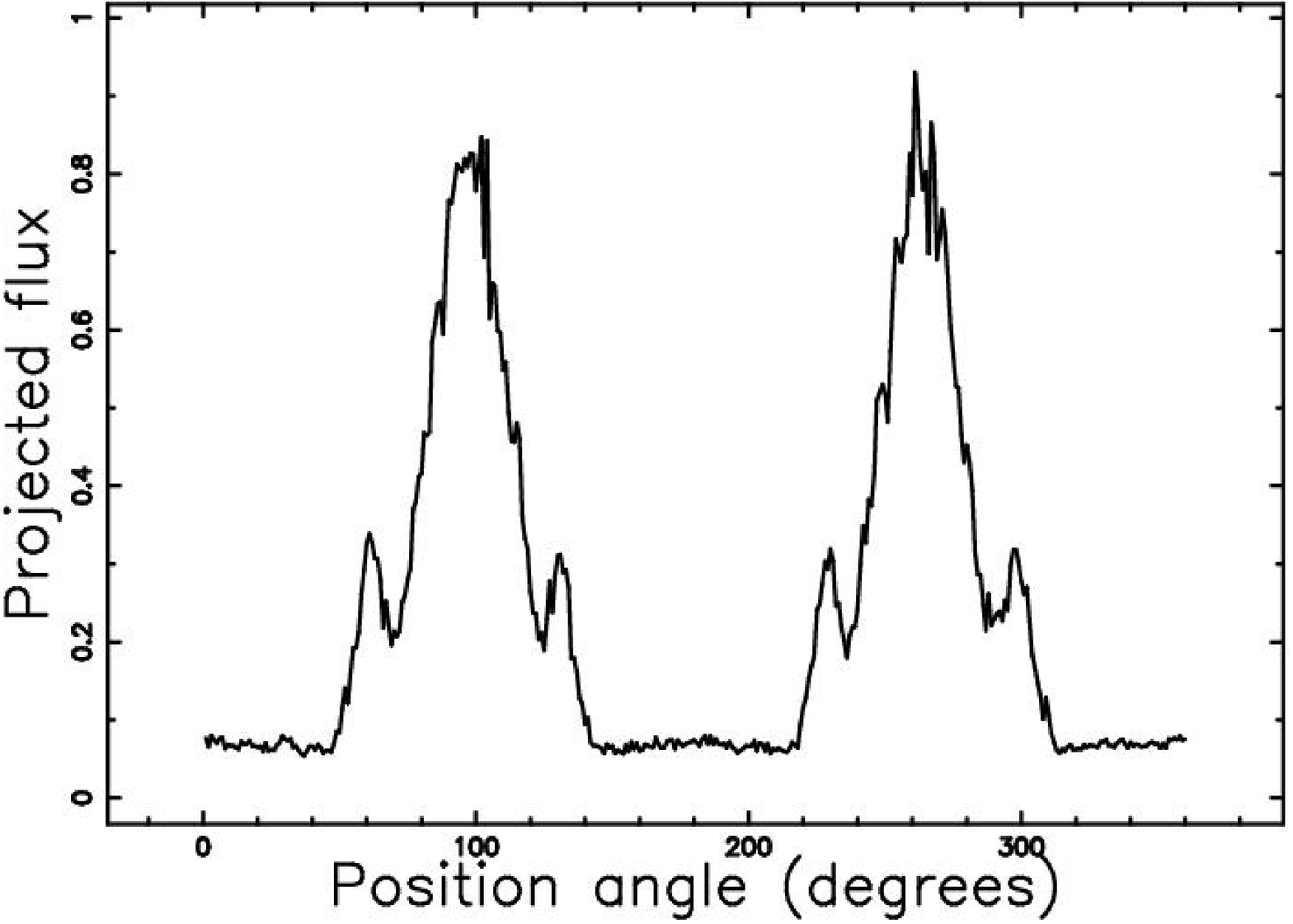}
\end {center}
\caption
{
Azimuthal maximum of the profiles of 
intensity as function of the position angle
in degrees.
Same  parameters as in Figure~\ref{f16}.
}
\label{f18}
    \end{figure*}

In Figure~\ref{f17}   
the asymmetry both in the peak to peak distance 
and the difference in the two maximum
is evident.
The ratio between the X-ray emission 
in the bright limbs (NE or SW) and toward the 
northwest or southeast  
( at 2 keV) is around 10 , see  Figure 5 top right
in 
       Rothenflug  et~al. (2004)
. Conversely  our theoretical  ratio , see 
Figure~\ref{f17}, is 9.84.
It is  also  possible  to plot the maximum of  the theoretical 
intensity as  function  of the position angle ,
see  Figure~\ref{f18}.
The reader can make a comparison
with the observational counterpart
represented by Figure~5 top right, dashed line in 
       Rothenflug  et~al. (2004)
.

\label{velocities}

\sectionb {7}{Conclusions}

The results of  astrophysical interest are now summarised:
\begin{enumerate}
\item 
A careful  analysis  of X-intensity of SNR  near the external 
regions predicts a complex behaviour as  given  by the three 
equations (~\ref{I_1}), (~\ref{I_2}) and  (~\ref{I_3}));
the fit  with exponentials as adopted in 
       Bamba  et~al. (2003)
, our equation~(\ref{exponential}),  is not supported by theoretical
arguments.
\item
The overall intensity profiles of SNR in the X-region,
(represented by the new formulae
(~\ref{I_1}), (~\ref{I_2}) and  (~\ref{I_3}))
show a characteristic "U" shape  that can be explained 
by the mathematical 3D diffusion.
The predicted  ratio  between the
intensity in the ring region and in the center region (7.15) 
is in agreement with the observations (5.4 -10.3).
\item
The  agreement between observed and  theoretical
overall "U" profiles allow us  to exclude
the presence of self-absorption (optically thick case),
see simulation in Sect.~6.2.
\item
The Monte Carlo 1D random walk
with length of the step as given by the Bohm diffusion 
 provides the connection 
with the physical processes as well a new 
formula for the  damping length~(\ref{lrwnu}),
for the mean square displacement~(\ref{r2pc}) and 
for the diffusion coefficient in the presence of drift
(\ref{formulaud}).
The mean square displacement , equation~(\ref{r2pc}),  turns out to be
independent  from the selected band of synchrotron emission .
This independence explains the morphological  similarities 
between the various regions of the e.m. spectrum in SNRs.

\item
The SNR  analysed here ,SN1006, presents  a spatial asymmetry 
at  the  northwest-southeast axis as well as a corresponding
asymmetry in the flux.
They can both be simulated once the concepts of 
diffusion from many injection points and variable flux of velocity are 
introduced, see Sect.~6.6.
\end {enumerate}

The asymmetric random walk  on an unbounded lattice in 1D once 
 coupled
with the concept of damping length can provide :
\begin{itemize}
\item  a new way to deduce the magnetic field  , $H_{-4}$=1.2  ,
       against   $H_{-4}$=0.15  given by the equipartition arguments
\item  a reasonable explanation  for  the concentrations 
       of relativistic electrons 
\item  a value for  the   ratio  between the
intensity in the ring region and in the center region (6.88) 
which  is in agreement with the observations (5.4 -10.3).
\end  {itemize}

The analysis in the gamma region can be carried out 
splitting in two the analysis : gamma rays from synchrotron
emission and gamma rays production in the interaction of CR 
with target material.

Eqs.  (\ref{postshock})-({\ref{formulaud}) are strictly applicable
only to non-relativistic plasma  and the corresponding
relativistic  equations based on the first analyses of the theory
of relativistic Brownian motion, can be considered a target for
the future, see
           Dunkel  \&  H{\"a}nggi (2005a)
     ,
           Dunkel  \&  H{\"a}nggi (2005b)
     ,
           Dunkel  \&  H{\"a}nggi (2006)
    and
           Dieckmann  et~al. (2006)
    .
A subject of further investigation is the role that the
acceleration of electrons and cosmic rays plays on the shock jump
conditions. According to~
       McKee (1987)
the key parameter is
\begin{equation}
w = \frac{p_{rs}}{p_s} \quad ,
\end{equation}
where $p_{rs} $ is the post-shock pressure in relativistic
particles  and $p_{s}$ is the post-shock pressure. The effect
of the flow of energy from the shock to the relativistic particles 
 is
to reduce the temperature in the post-shock region and to modify
the structure of the shock front.


ACKNOWLEDGEMENTS.  The numerical code  uses the subroutines extracted from
"NUMERICAL RECIPES"  for mathematical help.
The    plotting packages are    PGPLOT~5.2 developed by
T.J.Pearson  and 
PGXTAL developed by D.S. Sivia.

\appendix

\sectionb {Appendix A}{Particle acceleration}
\label{app_acceleration}
\label{timefermi}
The concept  of stochastic acceleration  started  with the
work by  
Fermi (1949)
in which it was  found that the collisions of a
 particle of energy E  with a cloud
produced an average gain in energy ($\Delta E$)
proportional to the second order in u/c (u cloud velocity
approaching that of light, c ):
\begin {equation}
\langle \frac {\Delta E}  {E } \rangle
=
\frac {8}{3} ( \frac {u} {c} )^2  \quad.
\end  {equation}
This  process is named  Fermi~II.

On introducing  an average length  of collision
$\lambda$,  the
formula becomes:
\begin  {equation}
\frac {d  E}  {dt }
=
\frac {E }  {\tau }   \quad,
\end {equation}
where
\begin {equation}
\tau  = \frac {4} {3 } ( \frac {u^2} {c^2 }) (\frac {c } {\lambda })
\quad.
\label {tau}
\end   {equation}
It should be remembered  that  with a mean free path between clouds
$\lambda$ the average time between collisions
with clouds is
\begin{equation}
2\frac{\lambda}{c}
\quad  ,
\label{twice}
\end{equation}
see for example  
       Lang (1999)
(pag.~467~)~.
This means that the particle reaches the cloud (our  lattice point)
in a time that is twice  that of  straight  motion;
it is therefore possible to speak  of pseudo-rectilinear
trajectories  between the scatterer--clouds.

The probability , P(t), that the particle remains in the reservoir
for a period greater than t is now introduced,

\begin {equation}
P(t) = e ^{- \frac {t} {T}}
\quad  ,
\end   {equation}

where T is the time of escape  from the considered region.
The hypothesis that
 energy is continuously
injected in the form of relativistic particles with energy
$E_0$ at the rate R, produces ( according to 
Burn  (1975))
the following probability density ,N,
which is  a  function of   the energy:
\begin {equation}
N (E)  =
\frac {R \tau} {E_0}
(\frac {E} {E_0})^{-\gamma}
\quad  ,
\label {eq:ne}
\end   {equation}
with
\begin {equation}
\gamma = 1 + \frac {\tau} {T }
\quad,
\end   {equation}
and $\tau$ is defined in (\ref{tau})~.
Equation~(\ref{eq:ne}) can be written as
\begin {equation}
N (E)  =K \;  E ^{-\gamma}
\quad  ,
\label {eq:neK}
\end   {equation}
where $K = \frac {R \tau} {{E_0}^{-\gamma +1}}$~.
A power law spectrum in the particle energy
has now been obtained;
it must be
 remembered  that  in this case
the ratio of acceleration to escape time  does not depend
on particle energy.

The strong shock accelerating mechanism  was later
introduced by  
Bell (1978a)        
and 
Bell (1978b)        
;
the energy  gain  relative to a particle that is crossing
the shock is:
\begin {equation}
\langle \frac {\Delta E } {E } \rangle =
\frac {4} {3} \frac {u} {c}
\quad.
\end   {equation}
This process is named  Fermi~I;
further on this process produces an energy spectrum of
 electrons
of the type:
\begin {equation}
N (E) dE \propto E^{-2} dE
\quad.
\end   {equation}
This  allows us to postulate a kind  of universal mechanism
that produces synchrotron radiation with intensity
I($\nu$)
$\propto$ $\nu^{-0.5 }$.

It is   interesting to point out  that the index $\gamma$
of the accelerated particles  depends on the parameters
of the shocks u and L , where L represents
the thickness and u the velocity of the shocks~.
In particular  when  $u~<~\frac {D}{L}$ where D is the spatial
diffusion coefficient  , the spectrum
of accelerated  particles becomes very soft,
see Figure~3 by 
       Ostrowski  \&  Schlickeiser (1996)
~.
The
quoted standard spectral index of -2  adopted here 
for the
electrons is calculated
without  escape boundaries
assuming that there is an infinite box;
       Ostrowski  \&  Schlickeiser (1996)
 added the
possibility of escape and showed its
effect on the solution.

\sectionb {Appendix B}{The adiabatic losses}
\label    {adiabatic}
Other types  of losses are those due to  adiabatic
cooling  ,see e.g.  
       Downes et~al. (2002)
,
which affects
all electron energies and not just the highest ones
as in   synchrotron cooling.
Following the approach  of
       Longair (1994)
 (~equation~(19.5)~), they are
\begin{equation}
-  \frac {1}{E}
  \bigl  (\frac{dE}{dt} \bigr )_{ad} =
  \bigl ( \frac {1}{R} \frac{dR}{dt} \bigr  )  =
   \frac {1}{t_{exp}}
\label{eq:adiabatic}
\end{equation}
\label{sec_adiabatic}
where R is the radius of the SNR and $t_{exp}$ is
the characteristic
time of  expansion.
The   adiabatic losses are   negligible
if  the time scale  of the expansion
$t_{exp}$  defined in
equation~(\ref{eq:adiabatic})
is  greater than    the time , $t_{rw}$~,
the relativistic particles have been in the emitting
shell:
\begin{equation}
t_{exp} \gg  t_{rw}
\label{eq:adiadis}
\quad  .
\end {equation}
In the case of SNR, the analytical solution
for the radius can be found in
       McCray  (1987)
~,equation~(10.27) :
\begin{equation}
R(t)= \bigl ( \frac {25 E_{expl} t_{SNR}^2 }
              {4 \pi \zeta_0 } \bigr )^{1/5}
\label{eq:radiussnr}
\quad,
\end{equation}
where $\zeta_0$  is the density of the surrounding  medium
which is supposed to be constant , $E_{expl}$ is the energy of
the explosion,
and  $t_{SNR}$ is   the age  of the SNR.
One should note  that this is a Sedov solution,
while SN~1006 is
young and not described by Sedov's formulae.
Self-similar solutions for
this stage of evolution were proposed
by~ Nadezhin (1981), Nadezhin (1985) and Chevalier (1982).
Upon inserting  equation~(\ref{eq:radiussnr}) in
equation~(\ref{eq:adiabatic}),
 the time scale of the expansion
takes the simple form
\begin{equation}
t_{exp} = \frac{5}{2}  t_{SNR}
\quad.
\label{eq:texp}
\end {equation}
Sect.~4.1   shows that
with our choice of parameters the inequality
given by equation~(\ref{eq:adiadis}) is always verified;
therefore the adiabatic losses can be neglected.

\References

\vskip2mm


\refb
Axford~ W.~I.,  Leer  E., Skadron G.   1978, in International Cosmic Ray
  Conference , p.  132

\refb
Ballet~J.   2006, Advances in Space Research , 37 , 1902

\refb
Bamba A.,  Yamazaki R.,  Ueno M.,    Koyama K.  2003, \apj, 589, 827

\refb 
Bell~ A.~R.  1978a, \mnras, 182, 147

\refb 
Bell A.~R.   1978b, \mnras, 182, 443

\refb 
Berezinskii V.~S.,  Bulanov S.~V.,  Dogiel V.~A.,     Ptuskin V.~S.
  1990, Astrophysics of cosmic rays, North-Holland, Amsterdam

\refb 
Berg H.~C.  1993, Random Walks in Biology, Princeton University Press,
  Princeton

\refb 
Blandford R.~D.,  Ostriker J.~P.   1978, \apjl, 221, L29

\refb 
Bohm D.,  Burhop E.,    Massey H.,  1949, in Characteristic of Electrical
  Discharges in Magnetic Fields, McGraw-Hill, New-York

\refb 
Burn B.~J.   1975, \aap, 45, 435

\refb 
Chevalier R.~A.   1982, \apj, 258, 790

\refb 
Crank J.   1979, Mathematics of Diffusion, Oxford University Press, Oxford

\refb
Dieckmann M.~E.,  O'C Drury L.,    Shukla P.~K.  2006, New Journal of
  Physics, 8, 40

\refb 
Downes T.~P.,  Duffy P.,    Komissarov S.~S.   2002, \mnras, 332, 144

\refb 
Drury L.~O.  1983, Reports of Progress in Physics, 46, 973

\refb 
Dunkel J.,  H\"anggi P.,  2005a, \pre, 71, 016124

\refb 
Dunkel J.,  H\"anggi P.,  2005b, \pre, 72, 036106

\refb 
Dunkel J.,  H\"anggi P.,  2006, \pre, 74, 051106

\refb 
Dyer K.~K.,  Reynolds S.~P.,    Borkowski K.~J.,  2004, \apj, 600, 752

\refb 
Ellison D.~C.,  Berezhko E.~G.,    Baring M.~G.,  2000, \apj, 540, 292

\refb 
Ellison D.~C.,  Reynolds S.~P.,  1991, \apj, 382, 242

\refb 
Ellison D.~C.,  Reynolds S.~P.,  Borkowski K. et al.
  1994, \pasp, 106, 780

\refb 
Ellison D.~C.,  Reynolds S.~P.,    Jones F.~C.   1990, \apj, 360, 702

\refb 
Ellison D.~C.,  Slane P.,    Gaensler B.~M.   2001, \apj, 563, 191

\refb 
Fermi E.   1949, Physical Review, 75, 1169

\refb 
Ferraro M.,  Zaninetti L.  2001, \pre, 64, 056107

\refb
Ferraro M.,  Zaninetti L.  2004, \physa, 338, 307

\refb 
Gould H.,  Tobochnik J.   1988, An introduction to computer simulation
  methods, Addison-Wesley, Reading, Menlo Park

\refb 
Gustafson K.~E.   1980, Partial Differential equations and Hilbert Space
  Methods, John Wiley and Sons, New York

\refb 
Hillas A.~M.  2005, \JPG, 31, 95

\refb 
Hjellming R.~M.   1988, Radio stars.
Galactic and Extragalactic Radio Astronomy, p. 381

\refb 
Jokipii J.~R.  1987, \apj, 313, 842

\refb 
Kirk J.~G.  1994, in Benz A.~O.,  Courvoisier T.~J.-L.,  eds, Saas-Fee
  Advanced Course 24: Plasma Astrophysics Particle Acceleration .
p.~225

\refb 
Klepach E.~G.,  Ptuskin V.~S.,    Zirakashvili V.~N.  2000,
  Astroparticle Physics, 13, 161

\refb
Krymskii G.~F.,  1977, Akademiia Nauk SSSR Doklady, 234, 1306

\refb
Lang K.~R.,  1999, Astrophysical formulae , Springer, New York

\refb
Longair M.~S.,  1994, High energy astrophysics, Cambridge University Press,2nd ed., Cambridge

\refb
McCray R. 1987  In: {\it  Spectroscopy of astrophysical plasmas}, 
eds.~Dalgarno A.,  Layzer D., 
Cambridge University Press, Cambridge

\refb
McKee C.~F.  1987  In: {\it  Spectroscopy of astrophysical plasmas}, 
eds.~Dalgarno A.,  Layzer D., 
Cambridge University Press, Cambridge
\refb
Morse P.~H.,  Feshbach H.  1953, Methods of Theoretical Physics, Mc
  Graw-Hill Book Company, New York

\refb
Nadezhin D.~K.  1981, Preprint ITEP-1

\refb
Nadezhin D.~K.  1985, \apss, 112, 225

\refb
Ostrowski M.,  Schlickeiser R.  1996, \solphys, 167, 381

\refb
Pacholczyk A.~G.  1970, Radio astrophysics. Nonthermal processes in
  galactic and extragalactic sources, Freeman, San Francisco

\refb
Parker E.~N.  1958, Physical Review, 109, 1328

\refb
Parker E.~N.  1965, \planss, 13, 9

\refb
Press W.~H.,  Teukolsky S.~A.,  Vetterling W.~T.,    Flannery B.~P.,
  1992, Numerical recipes in FORTRAN. The art of scientific computing,
  Cambridge University Press, Cambridge

\refb
Reynolds S.~P.   1998, \apj, 493, 375

\refb
Rothenflug R.,  Ballet J.,  Dubner G.,  Giacani E.,  Decourchelle A.,
     Ferrando P.  2004, \aap, 425, 121

\refb
Rybicki G.,   Lightman A.   1985, Radiative Processes in Astrophysics,
  Wiley-Interscience, New-York

\refb
Schlickeiser R.   2002, Cosmic ray astrophysics, Springer, Berlin

\refb
Skilling J.   1975, \mnras, 172, 557

\refb
Strom R.~G.  1988, \mnras, 230, 331

\refb
Vainio R.,  Schlickeiser R.  1998, \aap, 331, 793

\refb
V\"olk H.~J.,  Berezhko E.~G.,    Ksenofontov L.~T.   2005, \aap, 433,
  229

\refb
Wolfendale A.~W.   2003, \JPG, 29, 787

\refb
Zaninetti L.   2000, \aap, 356, 1023

\end {document}